\documentclass[10pt]{article}

\usepackage{anysize}

\usepackage{amsmath,amssymb,epsfig}
\def\one{{\,\hbox{1\kern-.8mm l}}}
\newcommand{\Dslash}{\not{\hbox{\kern-4pt $D$}}}
\newcommand{\pdslash}{\not{\hbox{\kern-2pt $\partial$}}}

\newcommand{\Comment}[1]{{}}
\newcommand{\tchi}{{\tilde \chi}}

\def\IZ{{\mathbb Z}}

\def\IR{{\mathbb R}}

\newcommand{\ra}{\rightarrow}

\newcommand{\bc}{\begin{center}}
\newcommand{\ec}{\end{center}}
\newcommand{\ba}{\begin{array}}
\newcommand{\ea}{\end{array}}
\newcommand{\beq}{\begin{equation}}
\newcommand{\eeq}{\end{equation}}
\newcommand{\bea}{\begin{eqnarray}}
\newcommand{\eea}{\end{eqnarray}}
\newcommand{\bmx}{\begin{pmatrix}}
\newcommand{\emx}{\end{pmatrix}}
\newcommand{\nn}{\nonumber}

\newcommand{\be}{\begin{equation}}
\newcommand{\ee}{\end{equation}}

\newcommand{\del}{\partial}
\newcommand{\half}{\frac{1}{2}}

\def\IB{\relax{\rm I\kern-.18em B}}
\def\IC{{\relax\hbox{\kern.3em{\cmss I}$\kern-.4em{\rm C}$}}}
\def\ID{\relax{\rm I\kern-.18em D}}
\def\IE{\relax{\rm I\kern-.18em E}}
\def\IF{\relax{\rm I\kern-.18em F}}
\def\II{\relax{\rm I\kern-.18em I}}
\def\IZ{\relax{\sf Z\kern-.35em Z}}
\def\Id{\relax{1\kern-.32em 1}}
\def\IG{\relax\hbox{$\inbar\kern-.3em{\rm G}$}}
\def\IR{\relax{\rm I\kern-.18em R}}

\normalsize

\title{\bf Developments in high energy theory}

\author{{\bf Sunil Mukhi}$^1$ and {\bf Probir Roy}$^2$\\[2mm]
{\it ${}^1$Tata Institute of Fundamental Research, Mumbai}\\
{\it ${}^2$DAE Raja Ramanna Fellow, Saha Institute 
of Nuclear Physics, Kolkata}}

\date{}

\begin{document}

\advance\hsize by -1cm
\advance\parskip by 6pt

\maketitle

\abstract{

This non-technical review article is aimed at readers with some
physics background, including beginning research students.  It provides
a panoramic view of the main theoretical developments in high energy
physics since its inception more than half a century ago, a period in
which experiments have spanned an enormous range of energies, theories
have been developed leading up to the Standard Model, and proposals --
including the radical paradigm of String Theory -- have been made
to go beyond the Standard Model.  The list of references provided here
is not intended to properly credit all original work but rather to
supply the reader with a few pointers to the literature, specifically
highlighting work done by Indian authors.}

\thispagestyle{empty}

\newpage

\tableofcontents

\section{Introduction}

High energy physics has been and continues to be the cutting edge of
the human scientific endeavour probing ever smaller distances below a
femtometer ($10^{-15}$ m). The objective of this effort is to unravel
the properties of the basic constituents of matter together with those
of the fundamental forces of nature. In the process, deep new ideas
concerning them are theoretically developed and experimentally
tested. See Ref.[1] for a brief historical review of the subject.

This subject emerged as an entity distinct from nuclear physics and
cosmic rays in the late 1950's with the laboratory availability of
intense and collimated beams of protons and electrons, accelerated
upto giga-electron-volt (GeV) energies. These beams were made to
impinge on nuclear targets leading to the production of new
particles. The latter could be observed using sophisticated detectors
and computer analysis. 

Such studies enabled rather precise measurements leading to many
discoveries of both new particles and new effects involving their
strong, weak and electromagnetic interactions. For more than three
decades since the mid-1970's, these results have been explained by
a theoretical framework known rather prosaically as the Standard
Model. Hundreds of its predictions have been verified with impressive
precision in dozens of experiments generating billions of data
points. 

High energy experimenters are currently engaged in scaled-up versions
of similar kinds of studies at tera-electronvolt (TeV) energies,
probing distances down to almost an attometer ($10^{-18}$ m). These
are nowadays conducted largely with colliding beams of particles and
antiparticles which annihilate or scatter. Also relevant here are
non-accelerator experiments (such as neutrino studies and searches for
proton decay or cosmological dark matter) at both overground and
underground locations, each involving a gigantic apparatus. 

In addition, this field has been a fertile ground for innovative, if
sometimes speculative, ideas trying to go beyond the Standard
Model. These have provided a rich kaleidoscope of theories, some of
them aimed at a final unification of all fundamental forces including
gravity. Attempts have been made to extend the reach of some of these
theories, based on an underlying string-theory picture, all the way to
the Planck energy scale $M_{Pl} = (8\pi G_N)^{-1/2}$, $G_N$ being
Newton's gravitational constant. $M_{Pl}$ is about $2 \times 10^{18}$
GeV in magnitude, and represents the energy scale at which the
gravitational interaction can no longer be treated classically and
quantum gravity necessarily comes into play.

A single unified theory describing all elementary particle
interactions is still an elusive goal. The Standard Model describes
the electromagnetic, weak and strong interactions, but only unifies
the first two.  Despite its spectacular
success in explaining a large set of available data, it has serious
inadequacies. Attempts to go beyond it have tried not only to address
those issues but also to unify all interactions. Furthermore,
utilising recently acquired facts from cosmological observations and
current ideas in cosmology [2], these efforts are aiming to provide a
coherent, if still incomplete, picture of the Universe. Quite a few of
these theories predict different kinds of distinct new signals in
particle collisions at multi-TeV energies.

This energy domain, sometimes called the tera-scale, is about to be
explored by an accelerator called the Large Hadron Collider (LHC) [3],
which will collide two proton beams head-on at a centre-of mass energy
of 14 TeV. According to the generally accepted cosmological scenario,
this divided by the Boltzmann constant was the order of the average
temperature of the rapidly expanding ball of radiation and particles
that was our Universe about one picosecond after the Big Bang. The LHC
aims to recreate such conditions in the laboratory. Located in Geneva,
Switzerland, and operated by CERN (European Organisation for Nuclear
Research), this machine has been constructed over the last several
years and is scheduled to start operation in late 2009. Data relevant
for this exploration are expected to come in within a year or
two. This prospect has infused researchers in the field with great
excitement and anticipation.

\begin{figure}[h]
\begin{center}
\includegraphics[height=5cm]{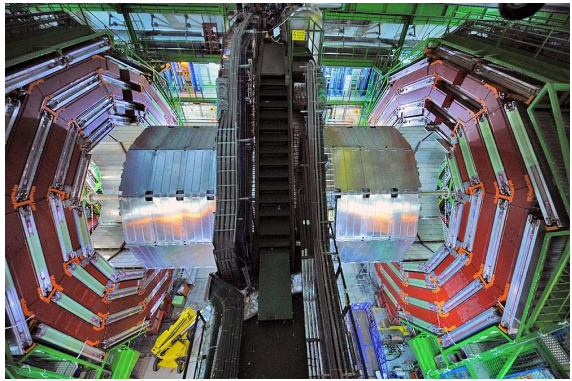}
\end{center}
\vspace*{-2mm}
\caption{The Compact Muon Solenoid (CMS) detector at the
LHC. Participating research groups in India are at BARC, Delhi
University, Panjab University and TIFR.}
\label{cms}
\end{figure}

Our purpose in this article is to provide a broad perspective on the
main theoretical developments that have taken place in high energy
physics and on the directions along which this area of research seeks
to go further. We have used some subjective judgement in selecting the
topics of our discussion. Admittedly, our coverage will by no means be
comprehensive. Nevertheless, we do hope to highlight the essential
strands. In spite of our focus on theoretical advances, we shall
mention key experiments which have triggered and sustained much of
that progress. Numbers quoted by us will always be in natural units $c
= \hbar = 1$ which are particularly convenient for describing
relativistic quantum phenomena.

\section{Rise of gauge theories}

\subsection{General aspects of gauge theories}

Looking back, one sees a revolutionary upsurge in the field during the
late nineteen sixties and early seventies. The crucial theoretical
development responsible for this was the rise of ``non-Abelian gauge
theories'', quantum field theories possessing an invariance under
continuous spacetime-dependent unitary transformations called ``gauge
transformations''. Such theories can exist only if they include spin-1
particles, whose existence is thereby an immediate prediction of the
theory. Non-Abelian gauge symmetry is a natural generalisation of the
well-known ``phase symmetry'' of field theories of electrically
charged particles i.e. electrodynamics, and leads to a rather
restrictive mathematical framework. It also bestows consistency on
relativistic field theories which are otherwise frequently
inconsistent due to negative-norm quantum states.

Gauge symmetry should correctly be thought of as a redundancy in the
degrees of freedom describing a quantum theory, with physically
observable degrees of freedom being gauge invariant. Without going
into the details of the gauge theory construction, we may mention some
general features. Gauge symmetry is defined by a Lie group, a
continuous group of transformations that is known in this context as
the ``gauge group''. An important Lie group is $U(N)$, the group of
unitary $N\times N$ matrices. This group has $N^2$ generators which do
not mutually commute. So as long as $N>1$, it is ``non-Abelian''. The
special case with $N=1$ is the group $U(1)$ of phase
transformations. This group is ``Abelian'', having only a single
generator. More general Abelian Lie groups are products of the form
$U(1)\times U(1)\times\cdots\times U(1)$, having many commuting
generators. A different Lie group, which occurs in the Standard Model
is $SU(N)$, a subgroup of $U(N)$ where the matrices are not only
unitary but also have determinant 1. This non-Abelian group has
$N^2-1$ generators.

A field theory with gauge symmetry contains ``matter fields'' that are
typically associated to spinless bosons or spin-$\half$ fermions. In
addition it contains the gauge fields referred to above, which have
spin 1. The number of gauge fields is precisely equal to the number of
generators of the group, specified above, and the transformation laws
for gauge fields are determined by the choice of gauge group. For the
matter fields we have more available choices, as they are specified by
irreducible representation of the Lie group, which can have varying
dimensionalities for the $SU(N)$ case. Having specified the
representation one can read off from the theory of Lie group
representations the precise manner in which the matter fields
transform. Interactions among gauge and matter fields are restricted
by the requirement that they be invariant under simultaneous gauge
transformations on all fields. This is a powerful constraint.

To summarise, the data required to specify a gauge theory is first of
all a gauge group which in our context can be restricted to contain
the factors: $$ U(1)\times U(1)\times
\cdots \times U(1)\times SU(N_1)\times SU(N_2)\times\cdots $$ together
with matter fields transforming in a specific representation of each
factor. As we will see in what follows, the Standard Model has the
gauge group $U(1)\times SU(2)\times SU(3)$, a rather special case of
the above structure. The representations for the matter fields are
somewhat subtle though, and have very important physical implications.

The idea of gauge symmetry is quite old and was originally mooted by
Weyl in 1929. His imposition on the Dirac free-electron Lagrangian of
an invariance under the transformations of a $U(1)$ gauge group
required the inclusion of a massless photon possessing a definite
interaction with the electron. Thereby one obtained the full form of
the (classical) action for relativistic electrodynamics. After about a
decade the formidable power of Weyl's idea was put to crucial use in
the consistent formulation of the quantum version of this theory,
Quantum Electrodynamics (QED). 

It should also be mentioned that just before the Second World War,
O. Klein had proposed a non-Abelian gauge theory to describe weak
interactions, but it had attracted little attention at the time. The
idea was independently revived in 1954 by Yang and Mills as well as
separately by Shaw. It was considered somewhat exotic through the 
following decade until it rapidly emerged as the correct formulation 
for particle physics in the late 1960's.

\subsection{Quantum Electrodynamics}

Quantum Electrodynamics was originally the theory of electron-photon
interactions. This was generalised to include the interactions of
photons with all electrically charged particles, there being one field
for each such particle (the field describes both particle and
anti-particle). 

Probability amplitudes for physical measurables appear in this theory 
as functions of kinematic variables and theoretical parameters. They are
computed as perturbation series expansions in powers of the fine
structure constant $\alpha = e^2/4\pi \sim (137)^{-1}$ where $e$ is
the electric charge. Although the lowest order results are generally
finite, higher-order terms contain integrals over momenta of ``virtual
particles''. As these momenta can go up to infinity, the corrections
end up being ultraviolet divergent.

At first it was thought that all higher order calculations would
therefore be impossible to carry out. However, this problem was
successfully tackled by Feynman, Schwinger, Tomonaga and Dyson. They
invented a procedure called renormalisation involving two basic
steps. First, the theory is ``regularised'' by cutting off its
high-momentum modes before computing anything. Second, the parameters
appearing in the Lagrangian, after including the regularised
corrections, are treated as ``bare'' (cutoff-dependent and unphysical)
quantities. One then defines ``renormalised'' parameters as functions
of these bare parameters in such a way that physical observables are
finite functions of the renormalised parameters to the given order in
perturbation theory. The process can then be extended order-by-order. 

The success of such a procedure is not guaranteed for all field
theories. Those for which it works are called
``renormalisable''. Otherwise, they are described as being non-renormalisable. 
The general wisdom in this respect today can be summed
up as follows. Non-renormalisable field theories can be used in the
approximate sense of an effective field theory over a rather limited
energy range. However, any fundamental field theory, valid over wide
energy scales, has to be renormalisable. 

Once renormalisation was successfully carried out in QED, physical
quantities -- such as the tiny splitting between the $2S_{\half}$ and
$2P_{\half}$ levels of the hydrogen atom (the ``Lamb shift''),
magnetic moments of the electron and the muon as well as measurable
scattering cross sections for many different physical processes --
could be calculated with high precision. Even more precise were
calculations of small energy shifts in positronium (an $e^+e^-$
bound state) and muonium (a similar bound state involving $\mu^+$ and
$e^-$) states and their decay rates. All these have been found to be
in incredibly precise agreement with experimental values, typically to
parts per million or even less. The pertinent point here is that this
procedure for QED cannot work without gauge invariance. The latter is
essential and plays a key role at every step of the renormalisation
programme, as amplified in the works of Salam and Ward.

\subsection{Non-abelian gauge theories}

As indicated above, following upon the success of the gauge principle
in QED, a field theory[4] with
non-Abelian gauge invariance was put forth. The original authors
utilised the gauge group of $2\times 2$ unitary matrices of unit 
determinant, called SU(2), but their treatment was actually extendable
to more general non-Abelian gauge groups. Importantly, they limited
themselves to the classical field theory and did not offer a way to
quantise it.

Despite the beautiful mathematics involved, one reason that theory
failed to attract much attention initially was the required presence
of three massless spin one bosons with non-Abelian charges. Such
particles were not found in nature. Adding a mass term for them to the
Lagrangian destroys the non-Abelian gauge invariance. Even with some
tweaking to render these particles massive, the best that could be
done was to identify them with three short-lived resonances called
$\rho$ (``rho'')-mesons (more on them later). But, as explained in the
next section, the $\rho$'s eventually turned out to be quark-antiquark
composites and not elementary spin-one particles. So the attempted
identification did not work. Another problem was that, though Yang-
Mills theories were suspected to be renormalisable, one did not know
how to actually show this in the absence of a detailed quantisation
procedure.

The situation changed dramatically after two great strides were taken
in utilising non-Abelian gauge theories. These led first to a
combined gauge theory of weak and electromagnetic interactions, called
the ``electro-weak theory'' and second, shortly thereafter, to a gauge
theory of strong interactions known as Quantum Chromodynamics
(QCD). Today, these two theories together comprise the Standard Model
of elementary particle interactions which will be the topic of
detailed discussion in our next three sections. While three of the
four known fundamental interactions are accounted for by the Standard
Model, the fourth (gravitation) is not a part of it. This is partly
because at the level of elementary particles the gravitational force
is many orders of magnitude weaker and can be safely ignored. Another
more fundamental reason, which will be discussed later on, is that the
Standard Model cannot be easily extended to include gravity.

The Standard Model has been found to be theoretically consistent
and has withstood all experimental tests directed at it for over three
decades. These tests have involved a variety of physical processes and
have made use of data generated from billions of high energy collision
events as well as many low-energy measurables. The Standard Model is
alive and well as of now, though its future may be uncertain, as
discussed below. It triumphantly proclaims today that we continue to
live in an age of gauge theories.

\section{Emergence of the Standard Model}

\subsection{A crisis and its resolution}

A crisis of sorts had engulfed high energy physics around the
mid-sixties. The existence of four fundamental interactions --
gravitational, electromagnetic, weak nuclear and strong nuclear -- had
already been established with a fairly decent idea of their relative
strengths. The first two were very well understood in classical terms
through the respective theories of Einstein and Maxwell. Moreover, the
second had been given a consistent quantum formulation in QED which
worked very well at least within a perturbative framework. However,
the knowledge that one had of the last two interactions was very
fragmentary and phenomenological and even the local field variables
in terms of which the theory should be formulated were unclear. 

All elementary matter particles, now as then, can be classified into
two categories: ``leptons'' (the electron $e$, muon $\mu$, tau $\tau$,
and the corresponding neutrinos $\nu_e,\nu_\mu,\nu_\tau$) which do not
experience strong interactions, and ``hadrons'' which do. Leptons are
always spin-half fermions, and are either stable or metastable. In the
latter case they decay via weak interactions, therefore relatively slowly,
having ``long'' lifetimes ranging from microseconds to picoseconds.

In contrast, hadrons can be either ``baryons'' (fermionic hadrons
carrying half-integral spins) or ``mesons'' (bosonic hadrons carrying
integral spins). Prominent among the baryons are the spin-half
nucleons, namely the proton, which appears so far to be
stable, and the neutron which has a long lifetime of slightly less
than 15 minutes. Also included in the list of baryons are the
meta-stable $\Lambda$ (lambda), $\Sigma$ (sigma) and $\Xi$ (xi, or
cascade) which carry an additional attribute or quantum number dubbed
``strangeness''. Among mesons the common ones are the spin zero $\pi$
(pion), $K$ (kaon), $\eta$ (eta) etc. which are all metastable. 

The list of hadrons also contains extremely short-lived (lifetime
$\sim 10^{-20}$ sec) particles such as the $\rho$ (rho), $\Delta$
(delta) and many others. More and more such particles were discovered
and soon they were over a hundred in number. This was a puzzling
situation. Given such a proliferation of short-lived hadronic
``resonances'', for a while the entire field-theoretic framework was
sought to be discarded in terms of a ``boot-strap'' theory of the
scattering matrix, or S-matrix\footnote{For some early work on this
from India, see [5].}. This formulation used some general principles
like analyticity and unitarity of the S-matrix to derive a few useful
equations [6] or sum-rules [7] for processes involving hadrons. But
that was as far as it went, and the theory did not turn out very
predictive.

As an alternative, the quark model [8] was advanced by Gell-Mann and
independently by Zweig. Quarks were proposed to be the bulding blocks
of hadrons. The proliferating hadrons were sought to be explained as
bound states of a few distinct quarks. Initially three flavours of
quarks were proposed, ``up'', ``down'' and ``strange''.
This proposal had some success in classifying the observed
hadrons, making it similar in some ways to the constituent theory of
the atom in terms of nucleons and electrons which explained the
periodic table of elements. However, beyond this success the quark
model made little progress and free quarks were not observed.

Towards the end of the 1960's, a path-breaking new `deep inelastic'
electron scattering experiment was performed at the Stanford linear
accelerator. In this experiment, electrons of energy $\sim 10$ GeV
were scattered highly inelastically, with a large invariant momentum
transfer, off protons and nuclear targets, producing a number of
hadrons in the final state. Though the produced hadrons were
unobserved, the scattered electron was detected and studied carefully.
This was a relativistic version of Rutherford's classic
`atom-splitting' experiment, more than half a century
earlier, studying the scattering of alpha-particles from a thin foil
of gold. 

Just as copious backward scattering in the latter established the
presence of a pointlike nucleus inside the atom, here also a
significant amount of backward scattering in the centre-of-mass frame
demonstrated [9] the existence of pointlike quarks inside the
nucleon. There was one major difference, however. Whereas a nucleus
could later be isolated by stripping an atom of all its electrons, all
attempts (even till today!) to isolate a single quark have proved
futile. Quarks seem to be perennially bound inside nucleons and more
generally within hadrons. Nevertheless, despite this overall
confinement, they seemed to behave nearly freely when interacting with
each other, as observed by a high momentum-transfer collision probe.

Additional data from other experiments, performed at GeV energies,
supported the above view. Highly inelastic scattering of neutrino
beams from nuclear targets, studied at the Proton Synchrotron at CERN,
Geneva, and the annihilation into hadrons of colliding electron and
positron beams, energised and stored in a number of storage rings,
yielded confirmatory evidence. Utilising the earlier quark model,
which had so far only been confirmed by hadronic mass
spectra, one was then led to\footnote{A comprehensive discussion of
the development of this paradigm may be found in [9].}  a new paradigm
for strong interactions, with three major inter-related aspects which
will now be described.

\subsection{Quark-gluon picture} 

The quark-gluon picture envisages hadrons as made up of spin-half
fractionally charged quarks bound together by spin-one uncharged
``gluons''.  Among
hadrons, the integral-spin mesons are quark-antiquark pairs while the
half-integral spin baryons are three-quark composites. For instance, a
nucleon consists of three quarks:
\bea
\hbox{proton} &=& \left(u_{\frac23}\,u_{\frac23}\,d_{-\frac13}\right)\nn\\
\hbox{neutron} &=& \left(u_{\frac23}\,d_{-\frac13}\,d_{-\frac13}\right)\nn
\eea
where $u$ and $d$ represent the up and down quark respectively, and
the subscripts represent the electric charges in units of the positron
charge. Other hadrons also involve the $s$ or strange quark, which explains
the origin of the strangeness quantum number referred to earlier. The quarks
listed above should be thought of as the basic constituents; they are
supplemented by clouds of quark-antiquark pairs and gluons that arise from
the vacuum in this strongly coupled theory.  In a high-momentum nucleon
these gluons, in fact, carry [10] about 50\% of the momentum.

While the original proposal of Gell-Mann and Zweig required only
quarks, gluons had to be added later on to fit the proposed framework
of a non-Abelian gauge theory, which necessarily has spin-1
quanta. Thus gluons are gauge particles and are the mediators of the
strong interactions. A key property of both quarks and gluons is
``colour'' as we now explain.

Even before the deep inelastic scattering experiments, a puzzle had
arisen for the quark hypothesis. There appeared to be a paradox
involving Fermi-Dirac statistics for identical quarks. Consider the
spin-$\frac32$ $\Delta^{++}$ resonance, consisting of three
up-quarks. As this is the lightest hadron with its quantum numbers,
the quarks must be in a spin-symmetric ground state. This would put
identical quarks in an identical state, violating Fermi-Dirac
statistics.

This was resolved by Greenberg and by Han and Nambu, who proposed that
each quark comes in three varieties called ``colours''\footnote{Like
``flavour'' before it, the term ``colour'' is merely a choice of name
for a particular type of attribute, or charge.}. This proposal solved the problem
because the three up quarks in a $\Delta^{++}$ can be antisymmetrised
in the colour degree of freedom, making the overall state
antisymmetric under the interchange of quarks, as required by
Fermi-Dirac statistics. Later studies with ``quarkonia''
(quark-antiquark bound states, analogous to positronium) as well as
direct hadro-production, bolstered this picture. Although the colour
hypothesis increased the total number of quarks by a factor of 3, a
disturbing lack of economy, it eventually triumphed and the strong
interactions are now understood to arise fundamentally from the
dynamics of colour.

While quarks come in three colours, the gluons which are exchanged
between them are labelled by a pair of colours; therefore one
expects nine of them. However it turns out that one of the nine
combinations is not required in the theory and therefore eventually
there are eight types of gluons.

Experimental studies also led to the discovery of additional
``flavours'' of quarks beyond up, down and strange, namely ``charm'',
``top'' and ``bottom''. The new quarks
have turned out to be heavier than the previous ones but experience
similar interactions. In this they are rather like the heavier leptons
(muon and tau) vis-a-vis the electron. 

\begin{figure}[h]
\begin{center}
\includegraphics[height=8cm]{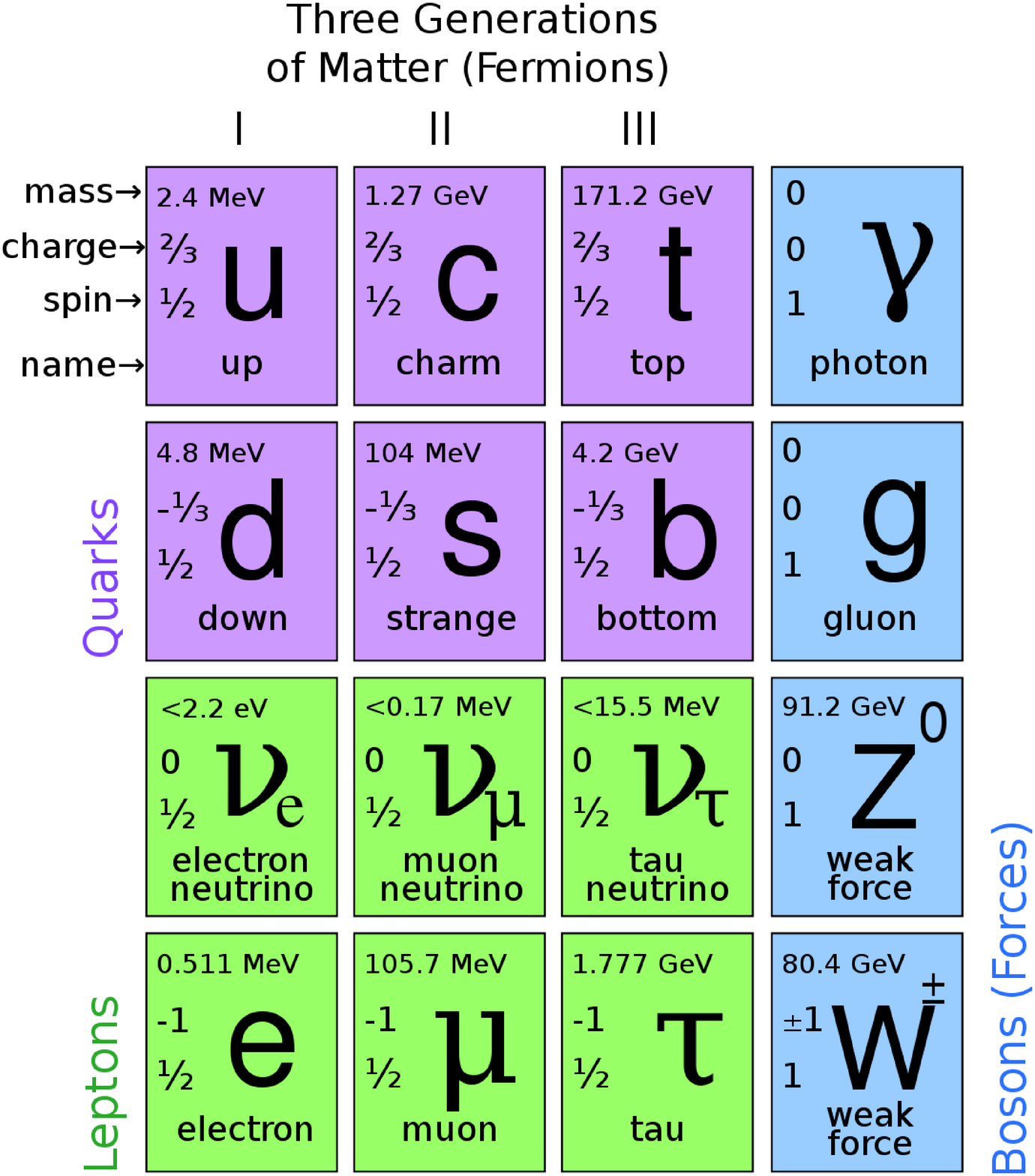}
\end{center}
\vspace*{-5mm}
\caption{A table of the Standard Model particles showing their
masses/mass bounds. {\it (From PBS NOVA, Fermilab.)}}
\label{standmod}
\end{figure}

In energy units, the electron mass is about $0.5$ MeV, while the up
and down quarks carry a few MeV of mass each. The muon and the strange
quark lie not far above 100 MeV in mass, while the tau as well as the
charm and bottom quarks lie in the mass range 1 - 5 GeV. The top quark
is the heaviest at nearly 170 GeV, just about the same as a gold
atom. This quark was discovered in the 1990's at the Tevatron
accelerator in Fermilab, U.S.A. Including the three neutrinos
($\nu_e,\nu_\mu,\nu_\tau$) which are taken in the Standard Model to be
massless, one ultimately has 6 leptons and 6 colour triplets of quarks
which are the building blocks for all observable matter particles,
stable or unstable, as depicted in Fig. \ref{standmod}.

\subsection{Asymptotic freedom} 

The Stanford experiment and other related investigations did more than
just demonstrate the existence of quarks inside a nucleon. They also
showed that, despite being bound, quarks behave almost freely when
studied with an external electromagnetic or weak-interaction probe
involving a large momentum transfer of the order of several
GeV. In effect, quarks interact feebly at small distances. 

As we saw above, soon thereafter a gauge theory was developed of
interacting quarks and gluons called Quantum Chromodynamics or QCD
based on the gauge group SU(3) of $3\times 3$ unitary matrices of unit
determinant. Let us focus on one important property of QCD. In any
field theory, the renormalised parameters discussed above turn out to
naturally depend on an energy scale (equivalently a distance scale).
They are said to ``run'' with the overall energy involved in the
process. The rate and direction of this running are determined by a
function of the coupling strength known as the
``$\beta$-function''. This function can be calculated in perturbation
theory for small enough values of the coupling strength\footnote{This
variation of coupling strengths with energy scale is often
referred to as a ``renormalisation group'' effect.}.

For most field theories (e.g. QED), such a calculation yields a
coupling strength that decreases at low energy (long distance) and
increases at high energy (short distance). This means that the
effective charge of the electron, measured at larger and larger
distances, appears smaller and smaller in magnitude: a phenomenon
known as screening. This is physically interpreted as being due to a
large number of virtual $e^+ e^-$ pairs in the photon cloud
surrounding the electron. Indeed, the electromagnetic fine structure
coupling alpha has the value $\sim 1/137$ mentioned above only when
measured in atomic processes involving a few electron-volt
energies. Because of screening, it rises to $\sim 1/128$ when
determined at the Large Electron Positron (LEP) ring at CERN at an
energy $\sim 90$ GeV.

The beta function for QCD was calculated by Gross and Wilczek [12] and
independently by Politzer [12]. It turns out to be negative, i.e. the
QCD coupling strength decreases as the probing energy increases or
equivalently we measure at shorter and shorter distances. At extremely short
distances the quarks therefore behave as though they were free. This
can be thought of as ``anti-screening'' behaviour, arising [9] from
the fact that the virtual gluon cloud around a quark has its own
self-interaction unlike the photon cloud surrounding an electron. 

This provides a rationale for the observed behaviour of the bound
quarks in deep inelastic processes. Moreover, a justification was
found for the successful use of perturbation theory in calculating the
corresponding observables, since at high energies one is in a
weak-coupling regime. Subsequent experiments have been able to measure
[13] the running or the energy dependence of the strong-interaction
coupling strength, called $\alpha_s$, agreeing within errors with the
theoretical calculation which now involves summing $\sim 50,000$ terms
at fourth order in perturbation theory. The calculation yields a
functional form of $\alpha_s$ which formally blows up at the energy
scale $\Lambda \sim 200$ MeV. The magnitude of $\Lambda$ turns out to
be exponentially related to the value of $\alpha_s$ at the Planck
energy mentioned earlier.

\subsection{Infrared slavery and confinement}        

Asymptotic freedom had a remarkable corollary which ultimately yielded
a decisive insight into the structure of hadrons. The perturbative
calculation of the QCD beta function, which indicates that the strong
coupling decreases at high energy, also implies that it increases at
low energy. Formally one finds that it goes to infinity beyond a
distance scale $\Lambda^{-1}$ of around a femtometer. While the
calculation itself ceases to be reliable at such scales (because 
perturbation theory breaks down at large coupling), it strongly
indicates that quarks are strongly coupled at large enough distances.

It was proposed by Weinberg that this phenomenon is the origin of
{\em quark confinement}. Though quarks behave almost freely at much
shorter distances, once the inter-quark separation becomes
significantly larger than a femtometer the effective force becomes a
constant (i.e. the potential rises linearly with distance) so that an
infinite amount of energy is required to separate two quarks.

An analogous phenomenon is observed in classical mechanics.  Two
objects joined by a rubber band experience little force when close by,
but the force becomes very strong when the rubber band is stretched.
While a field theory of quarks does not possess a fundamental ``rubber
band'', an analogy with superconductivity explains its possible
origin.  In superconducting systems, magnetic flux is confined into
long narrow tubes. By analogy, in QCD it is thought that a colour
(``chromoelectric'') flux tube stretches between quarks and binds them
together. If this flux tube has a constant tension, it would produce a
somewhat feeble force at short distances, but the energy residing in
it would grow with distance. This would imply that quarks rattle
around almost freely inside hadrons but are unable to come out, rather
like balls interconnected by rubber bands

There is now considerable support for the confinement phenomenon from
both theory and experiment, notably from numerical studies of QCD as
formulated on a discrete lattice. Nevertheless, colour confinement
remains a conjecture awaiting a rigorous proof. The precise statement
of this conjecture is that all objects carrying colour are permanently
confined (at zero temperature) and only colour-singlet bound states are
observed as physical states. In a high energy hadronic reaction,
coloured quarks and gluons inside the hadrons participate in an
underlying subprocess by scattering into other quarks and gluons. But
instead of being directly observed in the final state, the coloured
quarks and gluons ``hadronise'', which is to say that they extract
coloured partners from the vacuum to form colour-singlet hadrons.

A highly scattered quark or a gluon, or one emergent
from the decay of a very heavy parent, leaves behind its telltale
signature in the final state in the form of a spraylike jet of hadrons
travelling within a narrow cone around the direction of the parent
quark or gluon. The angle of such a cone would typically
be given by $\Lambda$ divided by the energy of the parent. Such hadronic
jets were detected during the 1970's and studied over many
years. They are among the important tools today in deciphering the dynamics
of the underlying quarks and gluons in high energy hadronic
processes. Their existence is one of many reasons why the quark
hypothesis is today considered to be firmly established despite
the absence of a rigorous proof of confinement.

The qualitative understanding of confinement described above also led
to a development of enormous significance. Instead of trying to derive
the existence of a confining flux tube from QCD field theory, one may
take an opposite point of view, as was done by Nambu and Susskind, and
propose that a flux tube of constant tension is a fundamental object,
namely a ``string''. In this proposal confinement is self-evident,
being built in to the theory, but the precise weak-coupling dynamics
of quarks at short distances then needs to be derived. This idea led
to the birth of an entire subject, String Theory, which has had
considerable impact on various fields of theoretical physics owing to
its highly consistent and powerful structure. It is described in some
detail later in this article.

\subsection{Weak interactions}

Slightly preceding the above exciting developments in the study of
strong interactions, came progress in the understanding of the weak
force. This is responsible for some kinds of nuclear radioactivity and for the
decays of several leptons and hadrons. The weakness of this force is
manifest in the relatively slow decay rate of these leptons and
hadrons, as compared with those that decay via strong interactions.

Until the mid-sixties, weak interactions were described by a
phenomenological Lagrangian field theory based on ideas due to Fermi,
Sudarshan, Marshak, Feynman and Gell-Mann. This theory used the notion
of ``currents'', which are certain field-theory operators made out of
fermion bilinears and can be interpreted as operators that create an
outgoing fermion out of an incoming one. The interaction in this
theory was chosen to be a product of two such currents, so that a pair
of incoming fermions instantaneously turned into a pair of outgoing
ones.  Thus it involved four fermion fields interacting at one
point. The strength of the interaction was defined by a dimensional
constant $\sim 10^{-5}\times\hbox{(proton mass )}^{-2}$ known as the
Fermi constant $G_F$.

Weak processes had been found to violate parity or reflection
symmetry, known to be respected by the other three fundamental
interactions. In relativistic field theory, parity violation is
allowed in theories where fermions have a definite ``chirality''. This
property tells us how the spin of the particle is correlated with the
direction of its motion, and is Lorentz invariant, therefore
well-defined in relativistic field theory. Fermions can be left-chiral
or right-chiral, while their anti-particles are correspondingly
right-chiral or left-chiral. In what follows, it will be important
that that mass terms normally mix opposite chiralities. Intuitively
this comes from the fact that a massive particle can be brought to
rest and then boosted in the opposite direction, reversing its
chirality. 

From nuclear beta-decay experiments, it became known that left-chiral
fermions (in some fixed convention) are the ones that participate
preferentially in weak interactions. This was evidence for parity
violation. Eventually it was found that right-chiral fermions never
emerge in these decays, so the parity violation is in fact maximal.
This fact was incorporated in the Fermi theory by using only
left-chiral fermions in the four-fermion coupling. Initially the
fermions participating in this interaction were assumed to be leptons
or hadrons (such as electrons or protons) but once quarks were
introduced it was natural to assign similar weak couplings for them
too. At a crude phenomenological level, this theory, with an extension
proposed by Cabibbo to include strange quarks, worked reasonably and
could fit a variety of experimental data on beta decay of nuclei as
well as free neutrons, as well as on the decays of metastable leptons
and hadrons.

The trouble, however, was that the theory was non-renormalisable. This
meant that higher-order computations were impossible to perform. Thus
the Fermi theory could not possibly be a fundamental theory of the
weak interactions and a new idea was urgently needed. One such idea
was put forward independently by Glashow, Salam and Weinberg
[14]. Their proposal, which incorporated some embryonic older ideas in
this direction, was that the four-fermion contact interaction should
be thought of as the effective version of an interaction that was
actually mediated by new bosonic particles. In this new theory, a
fermionic current would not interact directly with another one, but
rather would emit one of the new bosons which in turn would be
absorbed by the other current. At sufficiently low energies the result
could be mimicked by a four-fermion interaction but at higher energies
the presence of the new boson would significantly change the
theory. It was just conceivable that the new theory would be
renormalisable.

The new bosons needed to be massive and have one unit of spin. Three
such bosons were postulated, with the names $W^{\pm}$ and $Z^0$ where
the superscripts indicate their electric charges. Together with the
photon, which also has one unit of spin but is massless, they would
interact via a non-Abelian gauge theory whose gauge group would be
$SU(2) \times U(1)$. That this new theory of the weak interactions,
which we describe in more detail below, emerged successful speaks for
the amazing omnipresence of non-Abelian gauge theories in nature. It
bears no direct relation to the other non-Abelian gauge theory (QCD)
that we have already discussed and which describes the strong
interactions. In principle, either one of these theories could have
existed without the other one.

Returning to weak interactions, the $SU(2)$ gauge coupling strength
$g$ gets related to the Fermi constant $G_F$ through the mass of the
$W$-boson, via the equation 
$$ 
\frac{g^2}{8 M_W^2} = \frac{G_F}{\sqrt2}
$$ 
The successful features of the older Fermi theory are then completely
reproduced. However there is also a new and unprecedented consequence
of this proposal: the prediction of a new class of weak interactions,
mediated by the Z particle. This particle is electrically neutral and
is obliged to exist because of the Lie group structure of non-Abelian
gauge theory, and it leads to new interactions called ``weak neutral
currents'', which were discovered at CERN shortly thereafter in
neutrino scattering experiments.

The left- and right-chiral components of any fermion are
assigned to different representations of the $SU(2)\times U(1)$
gauge group in the electroweak theory. The assignment reflects
experimental inputs, notably parity violation. The left-chiral
electrons, which participate in weak charged-current interactions, are
chosen to be ``doublets'' of $SU(2)$, while the right-chiral ones are
chosen to be singlets. The same is done for the other charged
leptons, namely the muon and the tau. We
noted earlier that mass terms connect left and right chirality
fermions. Since all these fermions have long been known to have a
mass, it is clear that their right-chiral versions must be present in
the theory.

With neutrinos the situation is slightly different. They 
were long thought to be massless, therefore a logical possibility was
to simply have left-chiral neutrinos (in doublets of $SU(2)$, like the
charged leptons) and no right-chiral neutrinos at all. It is now known
that neutrinos have an extremely small mass, but because of its very
smallness we will temporarily neglect it and continue to discuss the
Standard Model in its original version with exactly massless
neutrinos. The issue of neutrino masses will be taken up later on. 

For the charged leptons we seem to be in a position to write a
consistent theory including their mass terms, since we have included
both left- and right-chiralities. However there remains a puzzle
involving the mass term. Because the two chiralities are in different
representations of $SU(2)$, the mass term connecting them cannot be
gauge invariant. But gauge invariance is a key requirement for any
theory having spin-1 bosons to be consistent, so it simply cannot be
abandoned. The resolution, one of the cornerstones of the Standard
Model, is that one introduces a new scalar (spinless) particle. Next
one writes not a fermion mass term, but a ``Yukawa'' interaction term
coupling a left- and right-chiral charged lepton to a new scalar
particle, called the Higgs particle. Finally one chooses a potential
for the field $\phi$ of this particle that is minimised by giving
$\phi$ an expectation value $v$. This creates a mass term proportional
to the expectation value: $$
\phi {\bar \psi}\psi \to \langle \phi \rangle {\bar \psi}\psi 
= v {\bar \psi}\psi 
$$
By this mechanism a mass is generated for the charged leptons (and
similarly for the quarks). Remarkably, a property of gauge theory
assures that the spin-1 gauge fields also acquire a mass due to the
expectaton value of $\phi$. This is called the Higgs mechanism, and is
based on  an idea borrowed from condensed matter physics. 

Experiment determines the expectation value of the Higgs field to be:
$$
v=(2{\sqrt2}G_F)^{-\half} \sim 175 \hbox{ GeV}
$$ 
Remarkably the Higgs mechanism accomplishes the requirement of
``breaking'' gauge invariance in  an apparent sense -- producing
masses for chiral fermions and gauge fields that would naively be forbidden by
gauge invariance -- and yet retaining the same invariance at a
fundamental level thereby rendering the theory unitary and renormalisable,
i.e. consistent at the quantum level unlike its predecessor, the Fermi theory.

The Higgs field can therefore be described as the source of all masses
in the Standard Model. One consequence of this scheme is the predicted
occurrence of a spin zero particle, called the {\em Higgs boson}, with
couplings to the $W$ and $Z$ bosons and the charged fermions that are
proportional to their respective masses. This particle is yet to be
experimentally detected, but its presence is vital to the
theory. Although it was suspected that the electroweak theory would be
renormalisable, a brilliant demonstration of this fact [15] by 't
Hooft and Veltman was the final theoretical result that transformed a
speculative idea into a coherent and consistent one. Subsequently,
every essential feature of it, other than (as yet) the existence of
the Higgs boson, was confirmed by experiment.

Though they are all proportional to the Higgs expectation value $v$,
fermion masses cannot be calculated in the Standard Model. Rather, a
proportionality constant is included in each interaction with the
Higgs field to reproduce the observed mass. A striking feature of the
most general mass term incorporating all the quarks, was that the
symmetry called CP (the product of charge conjugation and parity)
would be preserved if there were just two generations of quarks,
namely the up and down quark (first generation) and the strange and
charmed quark (second generation). The work of Kobayashi and Maskawa
[16] showed that the existence of a {\it third} generation of quarks
could indeed lead to CP violation\footnote{CP is conserved by gauge
interactions, therefore it can be violated only by the Yukawa
interactions which are responsible for fermion masses.}. As this symmetry was
already known to be violated in the interactions of certain uncharged
``strange'' particles called $K$-mesons, a third generation of quarks
was thus theoretically required. Quantum consistency of the theory
further required a third pair of leptons, charged and neutral.

These leptons and quarks, the $\tau$ (``tau") and $\nu_\tau$ as well
as ``top'' and ``bottom'', were subsequently discovered.
Moreover, CP violation has been observed not only in ``strange''
mesons (those containing a strange quark) but also in
similar ``bottom-flavoured'' particles called B-mesons. The
observations are in precise quantitative agreement with the
theoretical prediction. 

In contrast to the fermions, the gauge bosons $W^{\pm}$ and $Z$ had
their masses predicted rather precisely in the electroweak theory to
be around 81 GeV and 90 GeV respectively (being the antiparticles of
each other, the $W^+$ and $W^-$ must have the same mass). These
predictions were beautifully confirmed when the three bosons were
discovered more than a decade after the prediction was made, in a
pioneering experiment performed at the CERN Super-Proton Synchrotron
(SPS). This discovery put the stamp of universal acceptance on the
electroweak theory. This was confirmed more strongly during the
nineteen nineties by later data from the LEP machine which generated
billions of events containing $Z$ bosons and and $W^+$-$W^-$ pairs.

Because of the direct product structure of the weak-interaction gauge
group $SU(2)\times U(1)$, the theory has two independent measurable
coupling strengths which in turn determine $\alpha$ (the
fine-structure constant) and $G_F$. Therefore one does not have a true
unification of weak and electromagnetic interactions.  Nevertheless,
the fact that the $W$- and $Z$-bosons and the photon are inextricably
combined in the theory does endow it with some sort of partial
unification. Due to its renormalisability, the electroweak theory
allows the calculation in principle of any observable to an arbitrary
order in the perturbation expansion in gauge coupling strengths.

So far we have only mentioned, without providing any details, that the
theoretical developments described above were subsequently
confirmed by experiment. Historically, by the mid-seventies both the
electroweak theory and QCD had been formulated as consistent quantum
field-theoretic models of the electromagnetic, weak and strong
interactions but had not yet been extensively tested.  At this stage
particle theorists could be said to have presented a definite challenge
to their experimental colleagues. Would they be able to devise experiments
to exhaustively test this impressive theoretical edifice?  And if so,
would it survive or be demolished? In the next section we discuss how
experiments to date have tested the theory.

\section{Present status of the Standard Model}

\subsection{Electroweak theory}

Detailed predictions of the electroweak theory may be (and have been)
made using the tool of perturbation theory in the two dimensionless
electroweak coupling strengths. The values of these two couplings are
much smaller than unity, therefore the perturbative framework is
excellent for computing measurable quantities order by order and
comparing those with experimental numbers.

\begin{figure}[h]
\begin{center}
\includegraphics[height=11cm]{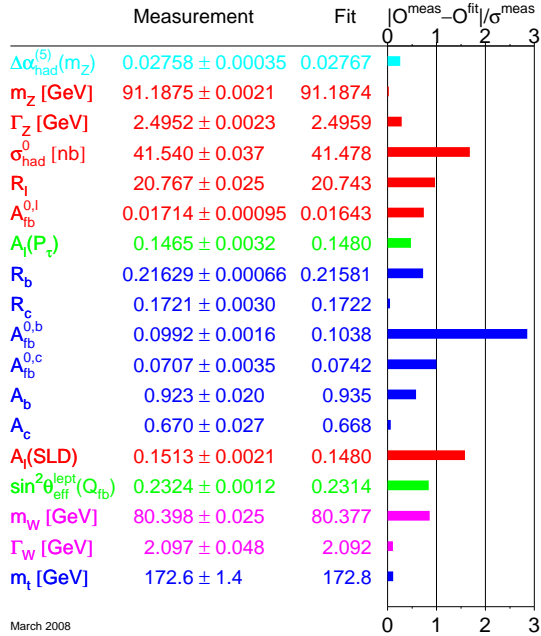}
\end{center}
\vspace*{-22mm}
\caption{Precision tests of the electroweak theory. {\it (From
G. Altarelli, arXiv:0805.1992.)}} 
\label{pullplot}
\end{figure}

Instead of providing detailed definitions of the measured quantities
and describing the procedures for their measurement, we exhibit a
diagram that illustrates the accuracy of comparison between theory and
experiment. Fig. \ref{pullplot} is a recent [17] ``pull-plot'' for eighteen
electroweak quantities, measured at different accelerators, but most
accurately at the Large Electron Positron (LEP) storage ring at CERN.
Here the ``pull'' for any measured quantity is defined as the absolute
value of the difference between the mean measured and the theoretically
fitted numbers, divided by the standard deviation in the measurement. 

\begin{figure}[h]
\begin{center}
\includegraphics[height=8cm]{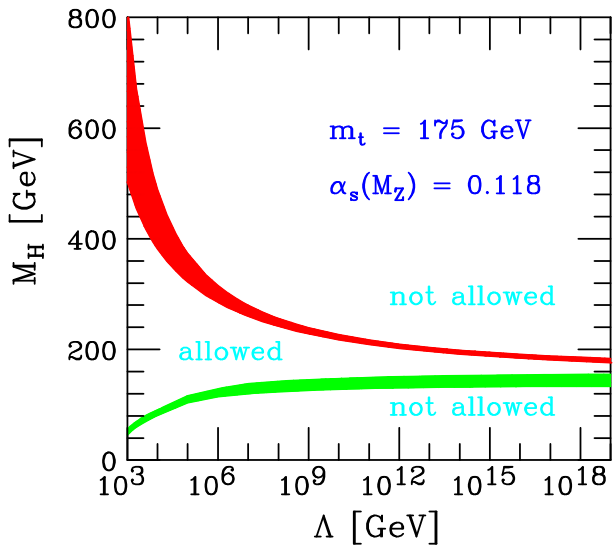}
\end{center}
\vspace*{-5mm}
\caption{Theoretical bounds on the Higgs mass as a function of the
scale $\Lambda$ upto which the Standard Model is assumed to be
valid. {\it (From A. Djouadi and R.M. Godbole, arXiv:0901.2030.)}}
\label{higgsbound}
\end{figure}

One can explicitly see from this plot, as well as from the actual
numbers, the very impressive agreement between the electroweak theory
and experiment. For many quantities, the agreement actually goes down
to the ``per-mil'' level (i.e. to one part in $10^3$) or better. Apart
from those listed in Fig. \ref{pullplot}, the ratio of the strengths
of the $ZW^+W^-$ and $\gamma W^+W^-$ couplings has also been found to
agree with the theoretical prediction at a percent level. The only
missing link in the electroweak theory now is the postulated but so
far unobserved Higgs boson. Indeed, one of the disappointing features
of this theory is its lack of prediction for the mass of the
Higgs. However, indirect arguments suggest [17] that the Higgs boson
lies somewhere between 114 and 190 GeV (see
Fig. \ref{higgsbound}). The lower bound is derived from the failure to
see the Higgs being produced at the LEP machine which discontinued
operations some years ago. The upper bound is a constraint from the
required agreement between theoretical calculations of electroweak
observables including the Higgs as a virtual state, and precision
measurements of the same. Interestingly, recent reports from the Tevatron
accelerator at Fermilab claim to have ruled out a Higgs boson in the
mass range 160--170 GeV at the 95\% confidence level.

The very characteristic coupling of the Higgs, being proportional to
the mass of the particle it couples to, is a useful tool in searching
for it. Using this fact, the Higgs is being sought ardently in ongoing
and forthcoming hadron collider experiments. If its mass is indeed in
the range mentioned earlier, it should be found without much
difficulty at the LHC, as clear from the calculation [18] of its
hadronic production cross section at the ``next-to-next-to-leading''
order. The quest for the Higgs has, in fact, the highest priority
among goals set out for the LHC machine. If the Higgs boson is not
found at the LHC even after a few years of running, a significant
feature of the electroweak theory will receive a death-blow.

In addition to the perturbative aspects of the electroweak theory,
spectacularly confirmed as described above, there are also physically
relevant non-perturbative aspects. A classical solitonlike solution to the
field equations of the theory, called a sphaleron, leads to physical effects
that are nonperturbative in the coupling constants. Sphalerons are
thought to play an important role in baryogenesis, i.e. the generation
in the early universe of the overwhelming excess of baryons over
anti-baryons that we observe in the cosmos today. Many popular models try
to calculate the measured baryon-to-photon ratio ($\sim 10^{-9}$) in
the Universe utilising sphalerons. However, with only one number and
quite a few models, the theory cannot be tested easily. Hence as of
now, all experimental checks of precisely measured electroweak
observables with theoretical calculations have been done within the
perturbative framework and the non-perturbative aspect of the electroweak
theory has so far not had any direct interface with experiment.

\subsection{Quantum Chromodynamics}

We turn now to the part of the Standard Model that describes strong
interactions, namely QCD. Here non-perturbative effects are more
significant since the coupling strength is large at or below the
theory's typical energy scale $\Lambda$. Sub-processes taking place
much above this scale can be accurately computed using perturbation
theory, but such a feat is much more difficult for any
non-perturbative part of a physical process. As the perturbative
computation applies at the quark-gluon level, while confinement causes
externally observed states to be colour-singlet hadrons, there is
really no experiment which tests the perturbative part alone.

Fortunately, QCD has an associated feature called factorisation which allows
one to separate rather cleanly the perturbative and non-perturbative
contributions to a physical process involving large transverse momenta.
Inspired by ideas from Feynman, Bjorken, Field, Altarelli, Parisi and others,
a methodology [19] for this has indeed been developed over the years. In this
procedure non-perturbative effects, involving the interface between
quarks/gluons on one hand and hadrons on the other, can be bunched
into quark and gluon ``distribution'' and ``fragmentation'' functions.
These are not calculable in perturbation theory, but can be argued to
depend on specific dimensionless kinematic variables, defined
vis-a-vis a specific hadron. These functions multiply the perturbative
parts of the relevant cross-section for a scattering process. In
addition, there are ``splitting functions'', describing quark-gluon
transitions, which can be computed in perturbative QCD. 

By cleverly combining various measurable quantities, one is able for
certain processes to isolate to a large extent the perturbative contribution
to a given process. Thus for this contribution, theory can be compared with data.
It has further been possible to parametrise the non-perturbative
distribution and fragmentation functions in terms of a few parameters
which can be fixed from the wealth of available data on hadronic
processes. Moreover, renormalisation group effects, mentioned earlier,
make these functions evolve with energy in a precise way predicted by
perturbative QCD; this behaviour has been [19] checked by comparison
with experiments performed at different energies. Researchers have
thus gained confidence in extrapolating those parametric functional
forms to much higher energies as will be probed at the LHC.

But there is a difficulty even in the domain of perturbative
effects. The magnitude of $\alpha_s$, the QCD coupling, is not so
small at the energies that are available today. Thanks to asymptotic
freedom, $\alpha_s$ does decrease to about 0.1 at an energy scale of
100 GeV, but it shoots up rapidly below that scale, becoming as large
as 0.35 at 1 GeV.  Thus in the multi-GeV regime where one would like
to compare with experiment, lowest order calculations are hopelessly
inadequate.  Therefore theoretical calculations have to take
subleading contributions into account. That still leaves higher-order
effects as well as non-perturbative effects, which together contribute
sizable uncertainties. It is therefore highly gratifying that the
result does match [13] the data within estimated uncertainties and 
experimental errors. There are many other instances of (quantitatively
less precise) agreement of theoretical computations using perturbative
QCD with experimental data at some tens of GeV. In summary it is fair
to say that there is broad, though not very precise, agreement between
perturbative QCD calculations and the relevant experimental data.

\begin{figure}[h]
\begin{center}
\includegraphics[height=9cm]{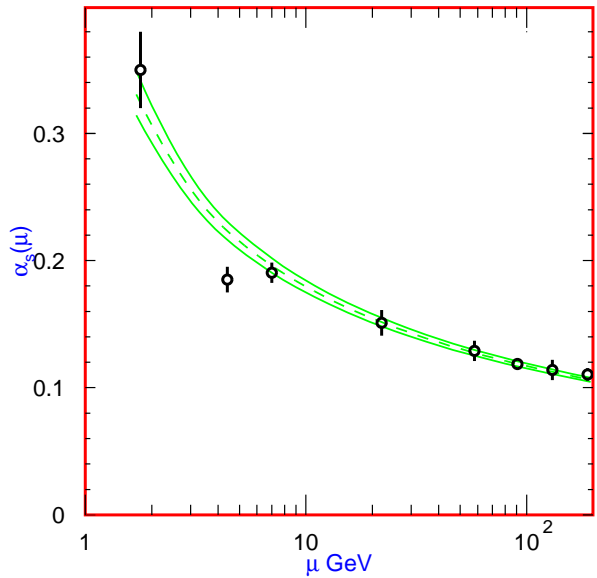}
\end{center}
\vspace*{-12mm}
\caption{Variation of $\alpha_s$ with energy. {\it (From
J. Iliopoulos, arXiv:0807.4841.)}} 
\end{figure}

Let us now come to non-perturbative aspects of QCD. Although
perturbation theory cannot be used, there exist other methods which
have had varying degrees of success, including the study of effective
theories, sum rules, potential models etc. But the most fundamental
non-perturbative approach to QCD is the discrete formulation called
lattice gauge theory [20]. Pioneered by K. Wilson in 1974, this
technique involves setting up a Euclidean version of QCD (i.e. with
imaginary time) on a discrete lattice in four dimensions. Quarks are
put on lattice sites and gluons treated as links between quarks at
adjacent sites. The lattice spacing $a$ plays the role of a regulator
with $a^{-1}$ being proportional to the high momentum cutoff while the
lattice size provides a practical infrared cutoff. Functional
integrals corresponding to various physical quantities are then
evaluated using Monte Carlo techniques which maintain gauge
invariance.

The idea is to calculate dimensionless combinations of the lattice
spacing with those quantities for smaller and smaller spacings,
eventually being able to make some statements in the limit when the
spacing vanishes. Initial difficulties in treating quarks dynamically
had led to the use of a ``quenched'' approximation in which quark
loops were ignored. But the availability of fast, dedicated
supercomputers as well as of more powerful algorithms and a better
understanding of cutoff effects, have enabled unquenched simulations
with quarks being treated dynamically. There are a variety of approaches
(Wilson fermions, Kogut-Susskind staggered fermions, Ginsparg-Wilson
fermions etc) in which successful attempts have been made towards
evaluating physical quantities.  One should emphasise here that
lattice QCD calculations have long overcome their initial quantitative
unreliability and today represent a rather precise science. To be
sure, there are errors -- due to finite lattice size, discretisation,
light quark-mass corrections etc. But these are much better understood
now and can be kept controllably small -- down to a few percent.

There have been basically two different types of lattice QCD
simulations, those at temperature $T=0$ and those at $T>0$. For the
former one now has a range of precise lattice simulations, many of
whose results are included in the tables of the Particle Data Group
[21]. With a rather small number of input parameters, these provide a
decent fit [22] to the observed hadron mass spectrum, including heavy
flavored mesons and ground plus excited states of different quarkonia
[23] as well as their level splittings. Another important milestone
has been the calculation [24] of $\alpha_s$ at the $Z$ mass pole near
90 GeV, which agrees to within a percent with the experimental value
from the LEP machine.

Among other successful calculations, mention can be made of the decay
constants of certain metastable mesons and form factors describing
strong interaction effects on the distributions of the final state
particles in those decays, as well as matrix elements in K- and
B-meson systems relevant to the study of CP violation and the
consequent extraction of the Cabibbo-Kobayashi-Maskawa mixing
parameters mentioned earlier. In addition, very useful calculations
have been performed of the distribution and fragmentation functions
that were discussed above. A spectacular success has been the
prediction [24] of the mass ($\sim 6.3$ GeV) of a $B_c$ meson, composed of
a bottom quark and a charm antiquark, which was experimentally
measured [25] a year later and agrees to within a percent.

We turn next to lattice simulations at finite [26] temperature. The
most important discovery in this type of investigation has been that
of deconfinement. We had mentioned earlier that at zero temperature,
the flux tube between coloured objects has roughly constant tension
corresponding to a linear potential at distances larger than a
femtometer. At $T>0$ the slope of the potential is found to decrease
until it vanishes at and above a critical temperature $T=T_c$. At
$T_c$, a phase transition to a state of colour deconfinement takes
place.

The order of this transition as well as the pertinent critical
parameters are found to depend sensitively on the chosen number of
quark flavours $N_f$ and on the masses of the quarks considered. We
know that the charm, bottom and top quarks are much heavier than the
rest. A reasonable procedure would be to assume that they decouple
from phenomena for which the relevant scale is a few hundred MeV. So one
can ignore them and take $N_f$ to be $2+1$, i.e. two light up and
down quarks and one heavier strange quark. Lattice simulations then
show [27] that the deconfining phase transition is second order, with
$T_c \sim 175$ MeV. 

An active area of research concerns the behaviour of the confining flux
tube with density. The confined and deconfined phases are found to
be separated by a crossover line at small densities, but by a critical
line at high densities. There is also indication of the presence at
high densities of a colour superconducting phase, with bosonic di-quarks
acting as Cooper pairs. Making these results more reliable and
quantitatively accurate is an important goal for those working on this
problem. It will significantly sharpen the interpretation of data from
current and future heavy-ion collision experiments discussed below.

One of the major arenas of the application of finite temperature QCD
is relativistic heavy-ion collisions [28] where the effects of
deconfinement can be studied and the formation (or not) of a proposed
quark-gluon plasma (QGP) phase [29] can be checked. Phenomena
associated with such collisions, are presently being studied at the
Relativistic Heavy Ion Collider (RHIC) at Brookhaven. These studies
will be extended to higher densities and energies at the LHC. RHIC
data already show the suppression of back-to-back correlations in jets
from central Au-Au collisions. This has been interpreted as due to the
formation of a hot and dense bubble of matter absorbing the jet that
crosses it. The produced hot matter shows strong collective
effects. This is evident from the observed elliptic [30] nature of the
flow of final-state hadrons. The latter is characteristic of a perfect
liquid with near zero viscosity, rather than a gas which would have
led to a spherical flow. Coupled with other related bits of evidence
from various patterns among the final-state hadrons, the following
broad view about the hot and dense object formed has taken shape. A
medium of coloured particles is surely being produced at RHIC with
high density and temperature; it then expands as a near-ideal
liquid. There is, however, serious doubt as to whether this medium is
truly the much sought after QGP. Hopes of definitive QGP formation and
its observation, for instance, through a clear suppression of heavy
quarkonium production, now lie with the ALICE detector at the LHC.

\section{Inadequacies of the Standard Model}

\subsection{Neutrinos}

Any impression from the above discussions that the Standard Model is
doing fine with respect to {\it all} experimentally studied phenomena
in high energy physics would be incorrect. Indeed, a chink in its
armour has already been found in neutrino physics. Recall that a
neutrino is weakly interacting and can travel quite a long distance in
matter before being absorbed. Neutrinos exist in three flavours:
electronic, muonic and tauonic. They do not have strong or
electromagnetic interactions, but couple to the $W^{\pm}$ and Z bosons
with gauge coupling strengths of the electro-weak theory, enabling
them to participate in weak processes. Therefore a neutrino is usually
produced in company with a charged lepton of the same flavour, in a
weak reaction mediated by $W^{\pm}$. It can be detected in a 
large-mass detector by any technique utilising the inverse of such a
reaction, namely an incident neutrino scattering from a target in the
detector and leading to the production of a charged lepton which can
be observed directly. 

It was mentioned earlier that neutrinos were assumed to be
massless and therefore right-chiral neutrinos were simply absent in
the Standard Model. But there is now indirect but quite
convincing experimental evidence that some of the neutrinos
do have masses. However those are extremely tiny (sub-eV) compared to the
masses of the other particles. As a result, some mechanism beyond the
Standard Model needs to be invoked to generate those masses.

\begin{figure}[ht]
\begin{center}
\includegraphics[width=12cm]{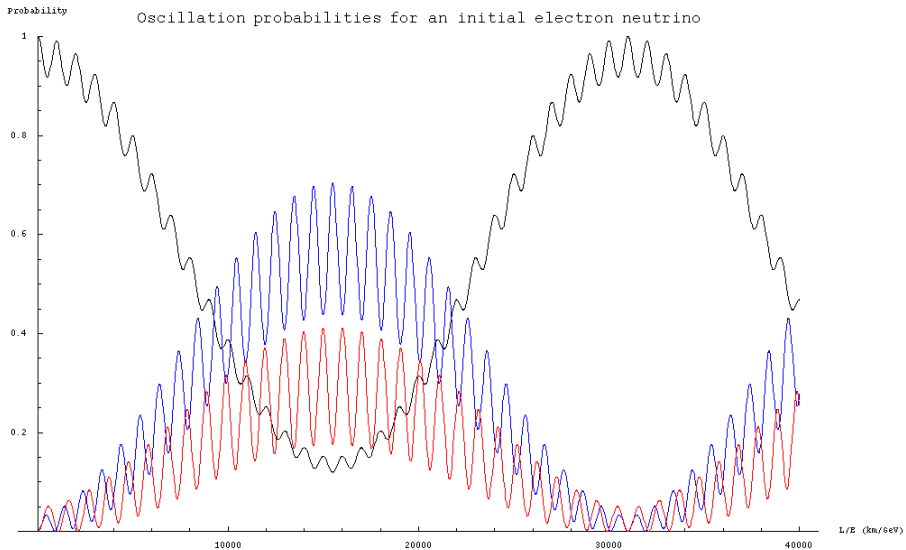}
\end{center}
\vspace*{-2mm}
\caption{Neutrino oscillations. The figure shows two neutrino flavour
waves in a beam, as well the resulting flavour oscillation probability.}
\label{oscillations}
\end{figure}

\begin{figure}
\begin{center}
\includegraphics[height=9cm]{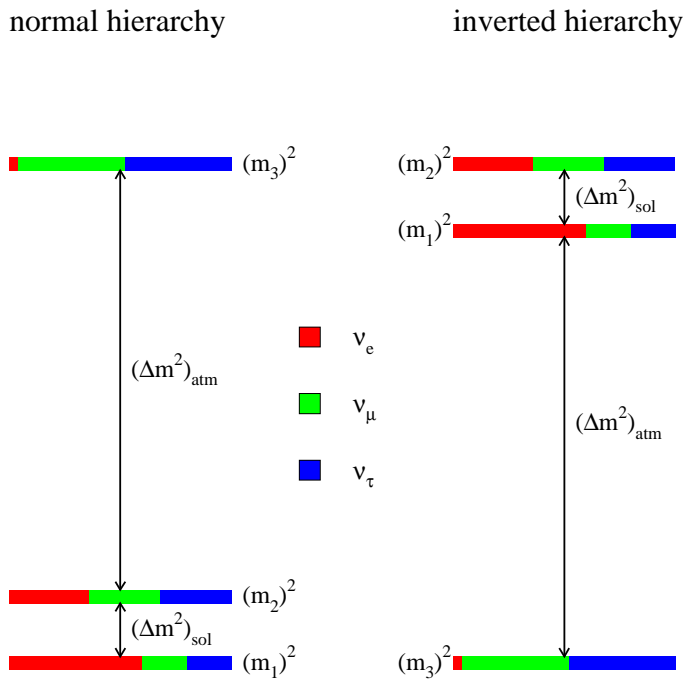}
\end{center}
\vspace*{-4mm}
\caption{Possible mass patterns of the three neutrinos. The left
(right) panel describes a normal (inverted) mass hierarchy with
$m_3^2$ greater than (less than) $m_1^2,m_2^2$. The shadings represent
the amounts of flavour mixing. {\it (From
A. deGouvea, arXiv:0902.4656.)}}
\label{mixing}
\end{figure}

It is found experimentally that neutrinos undergo oscillation based on
flavour conversion. The latter is a phenomenon in which a neutrino of
one flavour spontaneously converts itself into another of a different
flavour (and back) during free propagation
(Fig. \ref{oscillations}). This behaviour is known in field theory and
arises whenever eigenstates of mass are not eigenstates of flavour,
which in turn can happen when the mass terms involve ``mixing''. In
the neutrino sector there is substantial mixing, see
Fig. \ref{mixing}. Thus in a neutrino beam, the different flavour
components oscillate sinusoidally among themselve as a function of the
distance traversed divided by the beam energy. The function also
depends on the difference of squared masses of the two neutrinos
participating in the flavour oscillation. The experimental evidence
for flavour oscillation [30] comes from a variety of sources, namely
the sun, the atmosphere (where the neutrinos are decay products of
mesons produced by the aerial interactions of highly energetic primary
cosmic rays) as well as a number of nuclear reactors on the ground. On
analysing all these data, a consensus has emerged [31] today that at
least two of the three known neutrinos carry sub-eV masses.  Thus the
Standard Model has to be modified to include right-handed neutrinos
and a neutrino mass term has to be added to its Lagrangian.

At present only the differences in squared masses, $\Delta m^2$,
rather than the actual masses, of the neutrinos have been determined
by studying oscillations .  For one pair of neutrino flavours, solar
and reactor experiments fix $\Delta m^2$ to be about $8 \times 10^{-5}
\,{\rm eV}^2$ including a definite value for the sign. For another pair of
flavours, atmospheric neutrino studies measure $|\Delta m^2|$ to be
about $2.5 \times 10^{-3}\,{\rm eV}^2$ but they do not determine the
sign. These measurements enable us to deduce lower bounds of about
0.05 eV and 0.009 eV for the masses of two of the neutrinos.  Other
planned experiments on nuclear beta decays hope to measure different
combinations of the three neutrino masses. Meanwhile, all three neutrinos 
have been found to mix significantly
among themselves. Two of the three possible mixing angles are now
known reasonably accurately to be large ($\sim34^o$ and $\sim45^o$), while
the yet undetermined third angle is bounded above by $13^o$. 

Neutrinos, if stable, form the yet undiscovered ``hot dark matter'' in
the Universe. (here ``hot'' means having a relativistic Fermi-Dirac
velocity distribution). We shall have more to say on dark matter
below, but strong constraints do exist on hot dark matter from
cosmological observations. In particular, they imply an upper bound on
the sum of the masses of the three neutrinos. There is presently some
controversy about the precise value of this bound depending on which
set of cosmological data one chooses. However, an upper bound of about 0.6
eV on the sum of the three masses is widely accepted. 

Much remains to be found in the neutrino sector: their actual masses,
the mass ordering (i.e. whether the pair involved in solar neutrino
oscillations is more or less massive than the remaining state, see
Fig. \ref{mixing}), the value of the yet unmeasured mixing angle, more
definitive evidence of neutrino oscillation, more accurate values of
the mixing parameters etc. Another intriguing question is whether CP
is violated in the neutrino sector in the way it has been found to be
violated in the quark sector or otherwise. This issue may have
profound implications for the generation in the early universe of the
cosmological baryon asymmetry observed today. To unravel the several
remaining mysteries surrounding neutrinos, a number of experiments,
including one at the forthcoming India-based Neutrino Observatory
[32], have been planned and major new results are expected in a few
years.

Their tiny sub-eV values suggest that the masses of neutrinos have an
origin different from those of other elementary fermions, the quarks
and charged leptons. The latter masses are much bigger, ranging from
about half an MeV for the electron to nearly 170 GeV for the top
quark. Now the striking difference between neutrinos and all other
elementary fermions is that the former are electrically neutral while
the latter are all charged. For charged fermions, the only possible
mass term is one that links the left-chiral particle to an independent
right-chiral particle. However for a neutral particle such as the
neutrino, it is possible to write a mass term that links a left-chiral
particle to its own antiparticle (which is of course right-chiral). 
This is known as a ``Majorana mass'', and it violates lepton number
conservation since a particle with such a mass can spontaneously turn
into its own antiparticle.

Now because the Standard Model admits left-handed doublet fermions and
right-handed singlets, allowing Majorana masses means that neutrinos
in a single generation can have a $2\times 2$ symmetric matrix of mass
terms. The diagonal terms are Majorana masses which link the left- and
right-handed neutrinos with their own antiparticles, while the
off-diagonal one is the conventional ``Dirac'' mass term linking left-
and right-handed neutrinos to each other through the Higgs field:
$$
\begin{pmatrix}
m_{\hbox{\small Maj,L}} & m_{\hbox{\small D}}\\
m_{\hbox{\small D}}& m_{\hbox{\small Maj,R}}
\end{pmatrix}
$$

A particular structure for this mass matrix naturally produces, via a
``seesaw mechanism'' [33], Majorana neutrinos with tiny masses. This
mechanism works as follows. Arguments based on unified theories of
quarks and leptons, which we will discuss later, suggest that
$m_{\hbox{\small D}}$ is of the same order as the masses of other
elementary fermions, say a GeV. The right-chiral neutrino, being a
singlet of the electroweak gauge group, can have an arbitrarily large
Majorana mass, say $10^9$ GeV. Meanwhile the left-chiral neutrino,
being in an SU(2) doublet with a charged lepton, must have a 
vanishing Majorana mass. Thus the neutrino mass matrix
turns into: 
$$
\begin{pmatrix}
0 & m\\
m& M
\end{pmatrix}
$$ with $m\ll M$. Diagonalising this matrix gives us two physical
neutrinos with Majorana masses that are approximately $\frac{m^2}{M}$
and $M$. Thus one neutrino has a nonzero but extremely tiny mass that, with
our assumptions, naturally comes out to be sub-eV as desired. The other neutrino
is very heavy, with a mass $M\sim 10^9$ GeV.

Not only does this naturally provide ultralight neutrinos, but the
accompanying superheavy neutrinos have a desirable phenomenological
consequence. The decays of such heavy right-chiral Majorana neutrinos
could have triggered the generation of the cosmological lepton
asymmetry, leading eventually via sphalerons to the baryon asymmetry
observed in the Universe today. This is called baryogenesis via
leptogenesis [34]. In order to extend the seesaw mechanism to the
observed three flavors of neutrinos, each of the masses mentioned
above needs to be turned into a $3 \times 3$ matrix in flavour
space. The mechanism goes through, but it is still a major challenge
to reproduce the observed neutrino mixing pattern.

The elegance of the above seesaw mechanism notwithstanding, the issue of
the Majorana vs. Dirac nature of the neutrino needs to be addressed
directly. Indeed, a theoretical seesaw mechanism has been proposed [35]
for Dirac neutrinos also. Because massive Majorana neutrinos violate
lepton number conservation, the laboratory observation of a lepton
nonconserving process in which the neutrino plays a role would clinch
the issue in their favour. Such a ``gold-plated'' process is
neutrinoless nuclear double beta decay [36]. 

Ordinary beta decay of a nucleus involves an increase in the atomic
number (charge) with no increase in atomic mass, since it is due to
the process: $$ n\to p^+ + e^- +{\bar \nu}_e $$ Double beta decay
increases the atomic number by two units. In the normal version of
this decay, two neutrons in the parent nucleus simply decay
simultaneously. The final state then contains two electrons and two
(anti)-neutrinos. However, with Majorana masses
there is the possibility for these two anti-neutrinos to annihilate
each other (since the mass term converts an anti-neutrino back to a
neutrino). The final state then has two electrons {\em but no
neutrinos}. The observation of such a process
would be convincing evidence for Majorana neutrinos.

This proposed lepton number violating decay is extremely rare, with
a calculated half-life in excess of $10^{26}$ years, and indeed has not
so far been observed. Present limits, however, are close to values
expected from putative sub-eV neutrino Majorana masses. More probing
experiments [37] are currently under way with ${}^{76}{\rm Ge}$,
${}^{130}{\rm Te}$ and ${}^{100}{\rm Mo}$ nuclei, while there are
plans to use the nucleus ${}^{136}{\rm Xe}$ in a forthcoming
experiment. There is also an Indian proposal [38] to look for any
possible neutrinoless double beta decay of the ${}^{124}{\rm Sn}$ nucleus.

\subsection{The Flavour puzzle}

One of the puzzling features in the fermion spectrum of the Standard
Model is the fact that quarks and leptons come in three families or
``flavours''.  Each lepton family consists of a left-chiral doublet
containing a charged lepton and the corresponding neutrino, along with
corresponding right-chiral leptons that are singlets of
$SU(2)$. Similarly each quark family consists of a left-chiral doublet
containing an up-type and a down-type quark, and right-chiral singlets
of $SU(2)$, however we must remember that all quarks are also triplets
of the colour $SU(3)$ group.

Each family has identical gauge interactions. The only differences
seem to be in mass, the second family being heavier than the first and
the third heavier than the second. What could be the purpose of this
flavour multiplicity? The answer to this question clearly lies beyond
the Standard Model, which treats flavour in a mechanically repetitive
way. At the root of the puzzle presumably lies some new,
yet-undiscovered flavour symmetry. 

Of course, the flavour eigenstate fermionic fields have mixed Yukawa
couplings across families. As we have explained, fermion mass terms
arise from these couplings when the neutral Higgs field in them is
replaced by its vacuum expectation value. As a result, the fermion
mass terms involve $3 \times 3$ non-diagonal complex matrices
in flavour space -- leading to non-trivial mixing among the
differently flavoured fermions described by a unitary transformation.
The effects of quark mixing have been seen in transitions such as 
$$
s + \bar{d} \to \mu^+ + \mu^-, \quad b\to s + \gamma,\quad 
b\to s + \mu^+ + \mu^-
$$ 
observed in decays and mixings of heavy flavoured mesons. 

On the other hand, leptonic mixing has been studied in neutrino
oscillation experiments. As mentioned earlier, the observed mixing
patterns are very different for the quark and lepton sectors: a
feature not completely understood yet. However, one thing is quite
clear. The physical information in the $3 \times 3$ unitary matrix
which describes quark mixing is contained in three pairwise mixing
angles and one phase, all of which have been measured. In particular,
the phase, whose magnitude is in the ballpark of $45^o$, is
responsible for the violation of the CP symmetry which has been
observed in K- and B-mesonic systems in broad agreement with the
prediction of the Standard Model. As mentioned earlier, three is the
minimum number of families to have allowed such a phase. Having only
two families would have yielded a single mixing angle (the Cabibbo
angle) leading to a CP-invariant world. Now as observed by Sakharov,
CP violation is one of the requirements for baryogenesis, being
crucial to the formation of the presently observed baryon-asymmetric
Universe and hence life as we know it. Thus one can give an
anthropic argument for the existence of at least three families of
elementary fermions.

There has been much speculation [39] in the literature about the
nature of the flavour symmetry group which we call $G_f$. The idea
generally is to postulate some extra scalar fields, called ``flavon''
fields $\phi$, which transform non-trivially under $G_f$. The latter
is supposed to be spontaneously broken by a set of small dimensionless
parameters $\langle\Phi\rangle$ which can be interpreted as ratios
between vacuum expectation values of some flavon fields and a new
cutoff scale $\Lambda_f$. The Yukawa couplings $Y$ of the Standard
Model and the emergent fermion mass matrices $M_f$ can be thought of
as functions of $\langle\Phi\rangle $ and can be expanded as a power
series in it. 

In such an expansion, the lowest terms will yield just the Standard
Model expressions; but the next terms will involve dimension-six
operators scaled by two inverse powers of $\Lambda_f$. If $\Lambda_f$
is much larger than the electroweak scale, we shall practically not
see anything much beyond the Standard Model masses and
couplings. However, if there is new physics in the flavour direction
at a much closer energy scale $M$, lying say between 1 and 10 TeV,
these operators will show up in a variety of flavour-violating
processes such as radiative leptonic decays $$
\mu \to e + \gamma,\quad
\tau \to\mu + \gamma,\quad
\tau \to e + \gamma
$$ as well as in non-zero electric dipole moments of the electron and
the muon together with new contributions to their magnetic dipole
moments. New experimental quests to observe the said radiative
leptonic decays as well as any possible electric dipole moments of
charged leptons are being made continuously. Furthermore, there are
steady efforts to quantitatively improve our knowledge of the electron
and muon magnetic dipole moments. All these efforts hold the real
possibility of uncovering new flavour physics.

\subsection{The cosmological dark matter issue}

It is now known that nearly a quarter of the energy content of the
Universe is due to a form of distributed matter that has gravitational
interaction but is dark, i.e. non-luminous and electrically
neutral. This was inferred initially from the observed rotation rates
of galaxies which are faster than can be accounted for by the
gravitational pull of all visible matter in the sky. More recently,
sophisticated studies of both large-scale structure and temperature
anisotropies in the cosmic microwave background have confirmed this
conclusion. In fact, they have put it on a firmer and more
quantitative basis by being able to estimate the abundance of this
dark matter. 

The most natural explanation of the latter is the pervasive presence
of some new kind of stable or very long-lived particle(s) without
strong and electromagnetic interactions. (If dark matter had strong
interactions, that would have interfered unacceptably with the
nucleosynthesis process which took place within seconds of the Big
Bang in which the universe originated). Such a particle does not exist
in the Standard Model. Its presence would therefore require an
extension of the model. Other explanations, such as many massive
compact brown dwarfs or some different kind of clumped matter
distribution, have been disfavoured in the wake of gravitational
lensing studies [40a] on two merged galactic concentrations in the
Bullet Cluster, and the observed paucity of Massive Compact Halo
Objects (MACHOS) [40b]. The lensing studies have clearly resolved the
separation between the distribution of gravitational mass from that of
the luminous mass of those two merged galaxies.

Apart from determining the dark matter abundance, the above-mentioned
observations also tell us that dark matter is ``cold'', i.e. having a
non-relativistic Maxwellian velocity distribution. That fits in nicely
with the supposition that it is a ``Weakly Interacting Massive
Particle'' or WIMP. In many theories which go beyond the Standard
Model in the light of its other inadequacies, such a particle occurs
naturally -- typically in the mass range from a few tens to a few
hundreds of GeV. If it has weak interactions, it should be producable
(probably as a pair) at the LHC. It may quite possibly arise as one of the
products of the weak decays of some other heavier exotic particles
which are pair-produced from the proton-proton collision at the
LHC. 

However, once the dark matter particle is produced, it would escape
the detectors without any signal, rather like a neutrino. The major
difference from a neutrino would be the larger missing energy and
momenta in the experiment that would be carried off by a dark matter
particle. This in fact is the tell-tale signature of a WIMP at the LHC
and the planned International Linear Collider (ILC) [41] which aims to
collide $e^+ e^-$ beams at sub-TeV to TeV centre-of-mass energies. If dark
matter indeed consists of WIMPs, the latter are expected to be
whizzing around everywhere and more intensely in galactic halos because
of their mass. Experiments are currently in progress [42] trying to
detect these by scattering them via weak interactions from appropriate
target materials and studying the recoil effects. The expected cross
section is at a picobarn level. One group has even made a claim to
have seen such scattering events, but there has been no confirmation
in other searches. 

Another approach might be to detect high energy neutrinos or gamma
rays from the annihilation of two WIMPs which have come close to each
other at the centre of a massive star like the sun by its
gravitational pull. Continuous improvements are taking place in the
sensitivities of the detectors employed in world-wide search efforts
seeking WIMPs and new information is expected soon.

\subsection{Hierarchy problem}

The Higgs particle and its mechanism, needed to successfully generate
the observed masses of the weak vector bosons $W^{\pm}$ and $Z$, is
crucial to the electroweak sector of the Standard Model. As described
earlier, the Higgs potential is chosen to be minimised for a suitable nonzero
value of the field. This expectation value then provides masses to the
fermions and the gauge fields. In this process some of the original
Higgs field disappear to make up the extra polarisations of the massive
gauge bosons, but at least one Higgs field is left over. This too gets
its mass from its own vacuum expectation value.

At the classical level, the above proposal makes good sense.  But
problems start to arise in the Higgs sector when one tries to take
quantum loop corrections (and hence renormalisation) into account,
precisely what the model was constructed to enable.  Masses of
elementary scalar fields are unstable under quantum corrections, in
stark contrast to fermion masses. The latter are ``protected'' by an
approximate continuous symmetry called chiral symmetry.  This symmetry
ensures that fermions that are classically massless are also massless
in quantum theory, as a result of which if they are assigned small
masses classically (breaking chiral symmetry) then their quantum
masses will also be of the same order. This holds true even if the
theory is extended all the way to the Planck scale, the ultimate scale
beyond which we cannot ignore quantum gravity -- so fermions do not
acquire Planck scale masses upon renormalisation of the theory.

In contrast, the mass of the Higgs scalar in the Standard Model is
unprotected by any symmetry. The cutoff dependence of the quantum
correction to the scalar $\hbox{(mass)}^2$ is quadratic, with a
coefficient that is insensitive to that mass itself. Thus if one makes
the reasonable requirement that the quantum correction to the mass
should not be too large as compared with the original mass, one
is obliged to keep the cutoff within an order of magnitude of the
mass, which in the case of the electroweak theory is about a
TeV. 

Alternatively if we extend the theory to the Planck scale then the
natural value of the Higgs mass will be the Planck scale,
contradicting the fact that the Higgs mass is bounded by experimental
constraints within a range much below the Planck scale. Of course, in
a technical sense the renormalisation procedure can be chosen to fix
the physical Higgs mass to any value we like and place it in the range
where it is experimentally required to be. However this amounts to
``fine-tuning'' since in the cutoff theory the Higgs mass would be of
the order of the cutoff and one would then have to finely tune the
bare parameters, to roughly one part in $10^{17}$, to make the Higgs
mass desirably small in terms of renormalised parameters.  Moreover
this procedure would need to be repeated order by order in
perturbation theory, making the theory exceedingly ugly and unnatural
at the very least.

This problem of the Standard Model is called the ``naturalness'' or
``hierarchy'' problem. In fact, a principle of naturalness has been
enunciated [43] that, unless there is a symmetry that makes the
coefficient of the cutoff term vanish, it should generally be
nonzero. Since the expected Higgs mass is of the order of 100 GeV, the
hierarchy problem can be interpreted to say that the Standard Model is
only valid to within an order of magnitude of that, i.e. around a TeV
or so. That then is the energy scale at which the Standard Model would
be expected to break down with the emergence of new physics. This is
just the energy range that the LHC is going to probe.

\section{Physics beyond the Standard Model}

\subsection{Supersymmetry}

Supersymmetry is a proposed symmetry [44] transforming 
fermions into bosons and vice-versa. The fact that it changes the spin
of a particle makes it differ from the many other internal symmetries
occurring in particle physics in a number of ways. Since spin is
basically the eigenvalue of angular momentum, supersymmetry does not
commute with angular momentum. Instead, it transforms as a fermionic
operator carrying a spin of $\half\hbar$. Therefore it obeys {\em
anti-commutation} rather than commutation relations. Finally, it can
be shown that the anti-commutator of two supersymmetries when acting
on a field $\phi$ produces the spacetime derivative
$\del_\mu\phi$. This amounts to a translation. Indeed, supersymmetry
is the only ``internal'' symmetry that is intimately linked with
spacetime symmetry in this way.

There is no obvious experimental reason why supersymmetry has to be a
property of nature. For example no two elementary particles in the
Standard Model are, or potentially could be, related by
supersymmetry. This follows simply from the fact that all internal
quantum numbers, including the gauge group representations, of any two
superpartners must be identical (as supersymmetry commutes with other
internal symmetries). But in the Standard Model, the gauge particles
are in the adjoint representation while the fermions are all in
fundamental representations. That leaves only the Higgs particle as a
possible member of a supersymmetric pair, but it too can be shown not
to pair up with any of the known fermions.

Nevertheless, as we will explain below, there are pressing reasons why
supersymmetry is likely to be an approximate symmetry of nature. In this
scenario, the Standard Model must actually be {\em extended} by adding
a new, as yet undiscovered elementary particle for every known
particle. The ``superpartner'' of a known particle is sometimes called
a ``sparticle''. The superpartner of a spin-one gauge boson is a spin-half fermion 
called a ``gaugino'' (for example, the existing gluon and proposed
gluino would be superpartners). Similarly, each quark or lepton {\em of a
definite chirality} has a complex scalar as its superpartner. For
example the existing left-chiral electron would have as its
superpartner a ``left-selectron''. This terminology is admittedly
confusing, for the selectron -- being a scalar -- has no chirality; it
is merely the partner of a left-chiral particle.  

In this way the Standard Model can be extended into a Supersymmetric
Standard Model containing a number of sparticles. If supersymmetry were
exact, these would have had the same masses as the ordinary particles,
in evident contradiction with observation; if there were a spinless
``selectron'' with the same mass as the electron and similar
interactions as dictated by supersymmetry, it would surely have been
detected by now. So if supersymmetry is to be present we must further
postulate that supersymmetry is a ``broken symmetry'', which allows 
the masses of a pair of superpartners to be split. The splitting will be 
characterised by a mass scale (call it
$M_s$). So typically, in this hypothetical scenario, all the
yet-unobserved sparticles have masses of order $M_s$ larger than the
masses of the known particles and are likely to be discovered once our
accelerator energies are able to access that scale.

One can take issue with the supersymmetry idea on grounds of elegance:
why assume so many extra states? This, however, is an old game in
physics. Recall that Dirac postulated an antiparticle to every
particle so as to be able to reconcile special relativity with quantum
mechanics. Similarly rotational symmetry, and the consequent
properties of angular momentum in quantum mechanics, lead effectively
to a doubling of electron states as evidenced by the Stern-Gerlach
experiment.  Therefore it is not a priori absurd to accept that a
particular symmetry causes a doubling of the spectrum of particles by
invoking sparticles. The real question is: what does one gain by
postulating supersymmetry, and how can  the postulate be tested?

There are actually several advantages of a supersymmetric
scenario. Let us go through them one by one.

(1) Supersymmetry offers a satisfactory resolution [45] of the
naturalness or hierarchy problem of the Standard Model that was
discussed in a previous section. As we explained, the root cause of
this problem was that quantum loops induce a large correction to
the Higgs mass. However, in supersymmetric theories, for every quantum
loop contribution induced by a virtual bosonic particle, there is a
corresponding loop contribution of opposite sign induced by its
superpartner. Hence the dangerous correction gets cancelled and is absent.

Since supersymmetry, if present in the real world, is badly broken,
the argument has to be extended to cover that case. Certain types of
supersymmetry breaking terms in a theory (so-called ``soft'' terms)
preserve the good effects discussed above and continue to ``protect''
the Higgs boson from dangerous quantum corrections to its mass. In this way 
we are able to use supersymmetry to cure the hierarchy problem while at
the same time treating it as a broken symmetry in agreement with
observation.

This balancing act, however, works only upto a point. If the sparticle
masses are significantly more than an order of magnitude beyond the weak 
scale of $\sim100$ GeV, one begins to again encounter a naturalness problem.
This puts an upper bound on the extent of supersymmetry breaking one can
accept if the hierarchy problem is to be cured by it.

(2) Supersymmetry can dynamically trigger the Higgs mechanism which is
crucial to the generation of all masses in the Standard
Model. Choosing a potential with a minimum away from zero, so that the
Higgs field develops a non-zero expectation value, is rather {\em ad hoc} in
the Standard Model. Such is not the case if one has supersymmetric
extensions of the Standard Model, which require (at least) two complex
Higgs fields in doublet representations of $SU(2)$. One of these is
called $H_u$ because it generates masses for the up-type fermions,
while the other is called $H_d$ and does the same job for the
down-type fermions. 

Now the superpartner of the top quark (the heaviest quark), known as
the ``stop'', can via renormalisation group effects flip the sign of
the mass term in the potential for $H_u$ from positive in the
ultraviolet to negative in the infrared, turning it over at an energy
not far above 100 GeV. This then naturally leads to a nonzero minimum
of the Higgs potential generating the desired expectation value.
This feature of the supersymmetrized Standard
Model is called radiative electroweak symmetry breaking, and is hard
to arrange without supersymmetry.

(3) Supersymmetry provides a candidate for cosmological cold dark
matter. It is natural, though not compulsory, in supersymmetric models
to have a discrete symmetry called ``R-parity'' which is
exact. R-parity does not commute with supersymmetry, therefore it acts
differently on known particles and sparticles. Indeed while all
particles are even under it, every sparticle is odd. Conservation of R-parity
then implies that sparticles are produced and annihilate only in pairs.
Each produced sparticle will decay into lower mass sparticles and
particles, with the decay chain ending at the ``lightest sparticle''
(LSP). This particle, usually designated ${\tchi}$, has to be stable in
isolation as R-parity allows it to annihilate only in pairs.

The same processes would have happened in the early Universe, leading
to a substantial presence of these stable $\tchi$-particles which would
linger today as dark matter. One now knows the relic density of dark
matter quite accurately from cosmological observations. In most
supersymmetric models, the LSP is a neutral Majorana fermion which
does not have strong or electromagnetic interactions but does interact
via the weak interactions. That makes it an ideal cold dark matter
candidate. Two $\tchi$'s can annihilate via a virtual Z boson or a
virtual sfermion producing a matter fermion (quark or lepton) and its
antiparticle. This thermal-averaged cross section can be calculated
for the early Universe and from that the $\tchi$ relic density in the
cosmos can be derived. Assuming a $\tchi$ mass of about 100 GeV and the
weak couplings of the Supersymmetric Standard Model, the calculated
relic density turns out to be in remarkable agreement with that
deduced from cosmological observations\footnote{Interestingly,
other models -- not involving supersymmetry -- that possess weakly
interacting dark matter candidates, also appear to be consistent with
observations.}. 

(4) Supersymmetry provides a framework for the inclusion of gravity
along with the other interactions. We have already mentioned that two
supersymmetry transformations produce a translation. Now suppose one
wants to consider ``local'' supersymmetric transformations, i.e. those
in which the parameters are arbitrary functions of spacetime rather
than constants. Inevitably their anti-commutator will lead to local
translations, but these are just equivalent to general coordinate
transformations. It follows that local supersymmetry requires gravity,
and in a theoretical sense is {\em more fundamental} than gravity!

Local supersymmetry is often called ``supergravity'' and its fields
are the graviton field of spin 2 along with a superpartner of spin 3/2
called the ``gravitino''.  Since the graviton is necessarily massless
(corresponding to gravity being long-range), exact supersymmetry would
have rendered the gravitino massless as well. Fortunately the
supersymmetry breaking we have already invoked for other reasons gives
it a mass.

As far as we know, supergravity theories are not renormalisable and
therefore by themselves cannot provide an acceptable framework for
quantum gravity. However, Superstring Theory, which we discuss below,
is well-defined in the ultraviolet and gives rise to supergravity
coupled to matter fields as an effective field theory at energies
below the Planck scale.

(5) Supersymmetry leads to a high scale unification of the three gauge
coupling strengths associated with the gauge groups $SU(3)$, $SU(2)$
and $U(1)$ in the Standard Model.  We shall have more to say on this
issue later, but here we briefly
describe it only to highlight the role of supersymmetry. The
speculation has existed for some time that at a suitably high energy
the strong, weak and electromagnetic interactions may unify into a
single gauge theory with a simple gauge group and a single coupling
strength. This proposal is known as ``grand unification''. 

\begin{figure}
\begin{center}
\includegraphics[height=8cm]{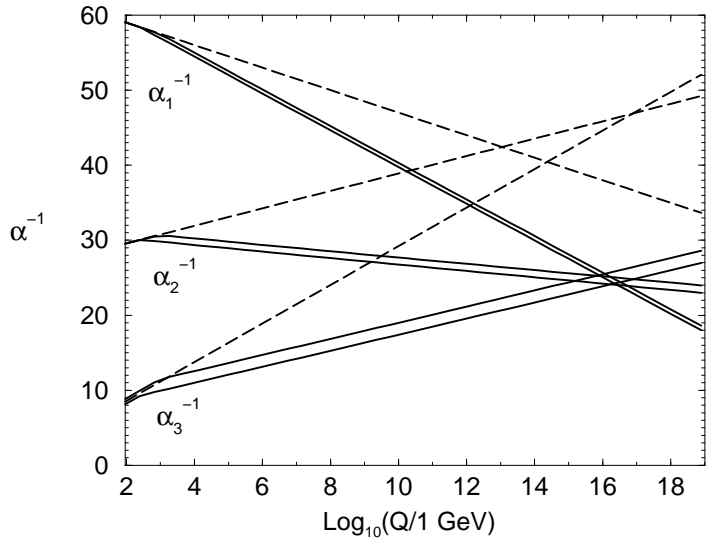}\\
\end{center}
\vspace*{-8mm}
\caption{The running of the $\alpha$'s in the non-supersymmetric (dashed lines) and 
        supersymmetric (continuous lines) cases. The double lines for the latter 
        indicate the uncertainty in the calculations. {\it (From
J. Iliopoulos, arXiv:0807.4841.)}}
\label{running}
\end{figure}

The viability of this proposal can actually be tested using known
experimental facts and the renormalisation group. We know the
magnitudes of the three relevant coupling strengths fairly accurately
at laboratory energies around 100 GeV. Their energy dependence, on
account of renormalisation group running, can be reliably calculated
in perturbation theory in the weak coupling regime which covers the
energy range from 100 GeV to the Planck scale. When this is done in
the Standard Model and the corresponding inverse fine structure
couplings $(\alpha_i)^{-1}$, $i = 1,2,3$ (cf. Fig. \ref{running}) are evolved to
high energies, it is found that the three curves do not intersect at a
single energy, as required by the idea of grand unification. When the
corresponding exercise is carried out in the supersymmetric extension
of the Standard Model, the curves are different because of
contributions from sparticles. Assuming the supermultiplet splitting
scale $M_S$ to be $\cal O$(TeV), the three curves now intersect
together (cf. Fig. \ref{running}) beautifully at a unification scale of about $2
\times 10^{16}$ GeV. Thus grand unification seems to work most
naturally in the presence of supersymmetry!

Having presented the various motivations for supersymmetry, let us now
discuss the simplest supersymmetric extension of the Standard
Model. This is known as the Minimal Supersymmetric Standard Model
(MSSM). Here the Higgs sector is extended, as explained earlier, from
one complex $SU(2)$ doublet to two, and these along with all the other
particles of the Standard Model are given superpartners. Moreover, the
conservation of R-parity is assumed. After working out the Higgs
mechanism, three of the eight real particles disappear (becoming
polarisation modes of the $W^\pm,Z$). That leaves five physical Higgs
particles: a pair of charged ones $H^{\pm}$ and three neutral ones
denoted $h$, $H$ and $A$. Of these the first two are even under the CP
transformation, while the third is odd.

The supermultiplet splitting scale $M_s$ is chosen to be in the range
of a TeV, though there is expected to be a spread in sparticle masses
and the LSP could well be as light as $100$ GeV. There are eight
gluinos of identical mass, matching the eight gluons. The physical
mass eigenstates are actually ``mixtures'' of electroweak gauginos and
higgsinos and correspond to two charge-conjugate pairs of
``charginos'' $\tchi^{\pm}_{1,2}$ and four neutral Majorana
``neutralinos'' $\tchi^0_{1,2,3,4}$, ordered in mass according to their
indices, with $\tchi^0_1$ being the likely LSP. 

As we have seen, every matter fermion has two chiral scalar
superpartners, e.g. left and right selectrons, left and right top
squarks and so on -- as we have seen (however if there is no right
chiral neutrino then the corresponding right sneutrino will also be
missing). Most of these sparticles and the Higgs bosons lie around
$M_s$, whatever that may be. As already mentioned, the LSP can be
quite a bit lighter. However, the lightest Higgs boson $h$ in the MSSM
turns out to have an upper mass bound around 135 GeV [44]. This is a
very important ``killing'' prediction for this model, since the LHC,
within its first few years, will clearly be able to confirm or
exclude the existence of such a particle. 

Attempts have been made to extend the minimal model by postulating
extra gauge-singlet fields. In this case the upper bound on $h$ gets
relaxed a bit, but it is difficult to push it much beyond 150 GeV and
impossible beyond 200 GeV. Thus it is fair to say that the entire idea
of supersymmetry in particle physics, broken around the weak scale,
will be ruled out if no Higgs scalar is found upto a mass of 200 GeV.

One of the unaesthetic features of the MSSM is the large number 
of arbitrary parameters in it. The Standard Model already 
has 18 parameters, while the MSSM, assuming no further input, has 106 extra 
parameters, 124 in all [46]. This is largely because the breaking of supersymmetry 
is introduced in the MSSM in an ad hoc phenomenological manner in
terms of the ``soft'' terms referred to earlier. These amount to giving unknown 
and different masses to all sparticles. However, there are more specific 
schemes of supersymmetry breaking involving either supergravity interactions 
or new gauge forces which can reduce the number of these parameters drastically. 

For instance, there is one such scheme with only three extra
parameters and one unknown sign. In these schemes, powerful
constraints ensue on the masses and mixing angles of many sparticles,
specifically on the mass ratios of the two lightest neutralinos and
the gluino. With sufficient data accumulated from the LHC and later
from the ILC, one may be able
to discriminate between such schemes of supersymmetry breaking. 

The most important experimental step awaited in regard to
supersymmetry, however, is the discovery of sparticles
themselves. Strongly interacting sparticles like squarks and gluinos,
if accessible in energy, should be copiously pair-produced (assuming
R-parity invariance) at the LHC. Each such sparticle will give rise to
a cascade of decay products eventually ending with the stable LSP. The
two LSPs will escape undetected carrying a large amount (more than a
hundred GeV) of energy which will cause a major mismatch in the
energy-momentum balance in the event. Such events will in general be
quite spectacular with multi-leptons and many hadronic jets as well as
large ``missing'' energy and momentum. They will not be accounted for
by Standard Model processes alone, though the elimination of the
Standard Model background will be a significant challenge. Such events
are eagerly anticipated at the LHC.

\subsection{Grand unification}

We briefly discussed the proposal that the three gauge couplings of
the Standard Model unify into one at a high scale. We also saw that
this is facilitated by (broken) supersymmetry. Let us now discuss the
primary motivation of grand unification. In the Standard Model, quarks
and leptons are necessarily placed in separate multiplets under the
gauge group. However, if all the quarks and leptons of each family
could be put in the {\em same} multiplet of some gauge group, the
resulting model would be extremely elegant and there would be just one
gauge force instead of three.

The grand unification idea was originally proposed by Pati and Salam
[47], though they did not have a simple gauge group with a single
gauge coupling. Georgi and Glashow [48] then proposed the simple group
$SU(5)$ and this leads to a grand-unified theory with a single gauge
coupling.  Historically, these theories were proposed without
supersymmetry, but we now know that supersymmetry is essential for
high-scale coupling unification, the cornerstone of this idea. So we
restrict our discussion to supersymmetric grand unification. 

$SU(5)$ is the simplest grand unifying gauge symmetry. Its
characteristic energy scale would be about $M_U \sim 10^{16} - 10^{17}$
GeV where it would possess a single gauge coupling with a unified
value of $(\alpha)^{-1}$ of about $1/24$,
cf. Fig. \ref{running}. $SU(5)$ then breaks into the Standard Model
gauge group $SU(3)\times SU(2)\times U(1)$ below that scale perhaps in
a manner analogous to the way that $SU(2)\times U(1)$ breaks down to
the gauge group for QED, namely $U(1)_{EM}$, at the electroweak
symmetry breaking scale near 100 GeV. Thus we have the chain: $$ SU(5)
\ra SU(3) \times SU(2) \times U(1) \ra U(1)_{EM} $$

The Standard Model has 15 chiral states per fermion family. The
counting is done as follows. There are two left-chiral quarks of three
colours each, as well as a left-chiral electron and neutrino, making a
total of 8 left-chiral particles. Except for the neutrino, each
particle has a right-chiral partner, making a total of 7. Thus we get
15 chiral states.  Now, among the low-dimensional representations of
$SU(5)$ we find the {\bf 5} and the {\bf 10}.
Thus it is natural to put these fifteen chiral states
into these representations\footnote{Technically they are put in the
{\bf 5} and the $\mathbf{\overline{10}}$.}. 

As a consequence one finds that the theory has heavy ($\sim M_U$) gauge
bosons denoted $X$ and $Y$ with respective electric charges 4/3 and
-1/3. These can convert quarks into anti-leptons and vice versa.
There are also extra heavy fractionally charged Higgs bosons with
similar properties. Both types of heavy bosons mediate baryon- and
lepton-number violating transitions. They thereby destroy the
stability of the proton, which can now decay into an anti-lepton and a
meson, albeit with a lifetime $\sim 10^{34}$ years which is much greater
than the age of the Universe $\sim 13.7 \times 10^9$
years. Experimentally, any kind of proton decay is yet to be observed,
but the current lower bound on the proton lifetime from the massive
water Cherenkov detector super-Kamiokande in Japan is $6.6 \times
10^{33}$ years. More sensitive experiments searching for proton decay
are on the anvil. Thus the supersymmetric $SU(5)$ theory is about to
face a major experimental challenge.

A grand unified theory based on $SU(5)$ has some important
inadequacies. Theoretically, it is not satisfactory that the chiral
fermions should be distributed in two different representations of the
gauge group instead of one. Experimentally, one finds it difficult
here to naturally accommodate a neutrino mass $\sim 0.05$ eV for which
we would like to invoke a right-chiral neutrino. In particular, in
this theory the seesaw mechanism, discussed above, would require a
heavy right-chiral neutrino that is an $SU(5)$ singlet -- an ugly
feature in a unified theory, for such a particle can receive a direct
mass bypassing the Higgs mechanism. Then the mass of such a particle can be
much heavier than the $SU(5)$ breaking scale $M_U$. 

For these reasons, another grand unified theory was put forward [49]
based on the orthogonal gauge group $SO(10)$. For an orthogonal group,
the most basic representation turns out to be its spinorial one. For
$SO(10)$, the spinorial representation is the {\bf 16}. This is ideal,
since the fifteen known chiral fermions can be accommodated and the
sixteenth member can be the right-chiral neutrino whose mass is
controlled by the unified scale at which $SO(10)$ breaks. Supersymmetric
$SO(10)$ grand unified theories have been worked out [50] in great
detail and can easily accommodate the known facts on neutrino
masses. They can also incorporate ideas like baryogenesis via
leptogenesis, with which the $SU(5)$ theory has problems. 

$SO(10)$-based grand unified theories also have specific predictions on
the lifetimes of various proton decay channels to be tested by
forthcoming experiments. However, this group has rank 5 unlike $SU(5)$
which has rank 4. Consequently, it accommodates a variety of symmetry
breaking chains to go down to the Standard Model gauge group and
admits more parameters. Hence there is greater lattitude in $SO(10)$
grand unified theories to increase the lifetimes of various proton
decay channels.

\begin{figure}
\begin{center}
\includegraphics[height=8cm]{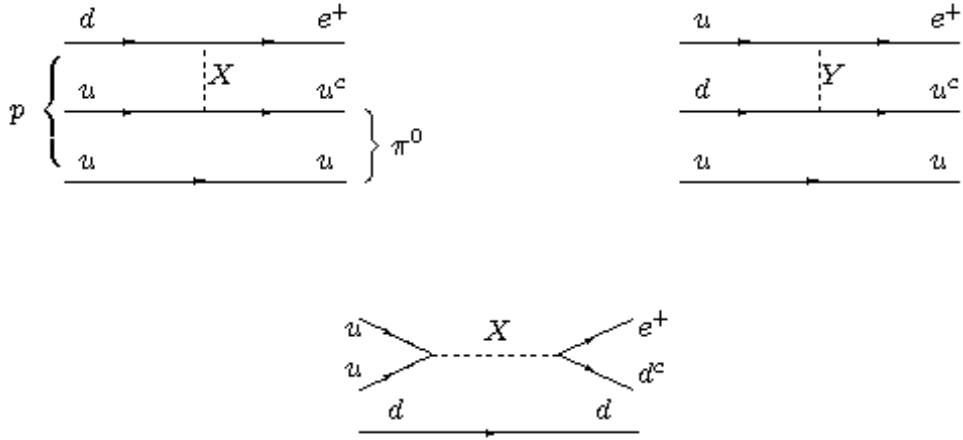}\\
\end{center}
\vspace*{-12mm}
\caption{Some diagrams for the decay of a proton into a positron and a neutral 
        pion. The superscript c on a quark means that it is the
charge-conjugate, or anti-quark.  {\it (From
J. Iliopoulos, arXiv:0807.4841.)}}
\label{protondecay}
\end{figure}

\subsection{Little Higgs theories}

A well-known phenomenon in quantum field theory, is that of the
spontaneous breakdown [9] of a global symmetry. This happens when the
Lagrangian is invariant under the symmetry transformations but the
vacuum is not. In consequence, there are degenerate vacua transforming
among themselves. These lead to massless bosonic modes called
Goldstone bosons. In case a small part of the Lagrangian also breaks
the symmetry explicitly, these states acquire small masses and then
are called pseudo-Goldstone bosons. The entire symmetry group, or only
a part of it, may be broken in this way. In the latter case a
particular subgroup of the original symmetry group remains
unbroken. 

Let us see how this notion arises in QCD at energies below or around 1
GeV. In this case one needs to consider only the two lightest quarks
($u$ and $d$), the other four being heavy enough that they can be
safely ignored.  Treating the light quarks as massless, there is a
global $SU(2)_L
\times SU(2)_R$ chiral symmetry under which the left-handed up and down
quarks form a doublet of the first factor and the right-handed up and
down quarks form a doublet of the second. Now this symmetry is not
seen (even approximately) in nature, but what we do see is the global
symmetry $SU(2)$ corresponding to isospin. This leads us to believe
that $SU(2)\times SU(2)$ is spontaneously broken [9] to the diagonal
subgroup $SU(2)_{L + R}$ around the energy scale $\Lambda_{QCD} \sim
200$ MeV. Since $SU(2)$ has 3 generators, the original symmetry group
has 6 generators of which 3 generators remain unbroken at the
end. Therefore, the other 3 must be spontaneously broken and
there should be a triplet of massless Goldstone bosons. If we now add
small mass terms in the Lagrangian of the theory, these bosons
will acquire masses and become pseudo-Goldstone bosons. These have
been identified with the observed triplet of pions.  This mechanism
then explains why the pions are relatively light.

The little Higgs idea [51] is to conceive of the Higgs scalar as a
pseudo-Goldstone boson of some broken symmetry and to generate its
mass in a way similar to the way the pion mass originates in QCD. Thus
we invent a global symmetry $G$ that spontaneously breaks down to $H$
at a certain scale (here, typically above a TeV). The pseudo-Goldstone
phenomenon produces a Higgs mass near the weak scale. It should be
noted that little Higgs models are really effective theories valid up
to $10-100$ TeV.

To understand why the Higgs has lost its famous quadratic dependence
on the cutoff, note that there are new particles in the model and they
occur in a certain pattern. In quantum computations, the contribution
from virtual particles of this kind cancels the dangerous cutoff
dependence of the Higgs mass. This cancellation is different from that
in the supersymmetric case where it occurs between virtual particles
of different spins. Here the cancellation takes place between virtual
particles of the same spin as a consequence of the special
``collective" pattern in which gauge and Yukawa couplings break the
global symmetries. The minimal scenario for which the scheme works
makes the choice $G = SU(5)$ and $H = SO(5)$ and is called the
``littlest Higgs'' model. In this case the new particles in the model
include a neutral weak boson $Z'$ and a top-like quark $t'$, in the
multi-TeV mass range.

\subsection{Extra dimensions}

An entire class of theoretical schemes going beyond the Standard Model
postulate the existence of additional spacetime dimensions beyond the
3+1 that we directly observe.  The original idea, due to Kaluza and
Klein and dating from the 1920's, was that there could exist extra
spatial dimensions that would not be in conflict with observation if
they were {\em compactified} i.e. defined on a compact manifold with a
very small volume. The simplest example would be a single dimension
valued on a circle of radius $R$.

The motivation for extra-dimensional  models is related to their role
in addressing the hierarchy problem, in unifying the fundamental
interactions (in  a way quite different from grand unification) and in
making physical sense of superstring theories that are consistent only
in higher dimensions. We will discuss all these motivations in what
follows. 

In a compactified theory, the volume $V$ of the compact dimensions
needs to be small enough to satisfy observational as well as
experimental constraints. The Kaluza-Klein idea was revived several
times, starting with the advent of supergravity and continuing into
the era of string theory. In the latter case, as we will see,
compactification is necessary to make contact with experiment since
the theory is formulated to start with in 10 spacetime
dimensions. String models employing such extra dimensions to predict
signals of physics beyond the Standard Model at laboratory energies
were constructed starting in the 1980's. In the last decade or so,
several problems with such models were overcome and they are now
capable of making quite specific predictions.

Though inspired by the string picture, these models can be discussed
in terms of their low-energy manifestation without referring to any
particular string theoretic scheme. Currently, there are two broad
categories of models in this genre: (i) those [51] with two or more
``large'' (sub-millimeter to sub-nanometer, or even smaller) extra
dimensions compactified on a torus, say, and (ii) those [52] with a
single ``warped'' small extra dimension (close to the Planck length
$M_{Pl}^{-1}
\sim 10^{-33} \hbox{cm}$). Here the term ``warp'' refers to the presence of a
metric having an exponential variation  along the extra dimension.

\subsubsection*{(i) Large Extra Dimensions}

If there exist extra spatial dimensions, the fundamental scale of
quantum gravity in the total spacetime, also called the
higher-dimensional Planck scale $M^*$, can be much smaller than the
4-dimensional Planck scale $M_{Pl}$. Now $M_{Pl}$ is
directly observable and known to be very large ($\sim 2 \times 10^{18}$ GeV),
and this largeness is directly related to the extreme weakness of the force of
gravity in our world. 

One is therefore tempted to argue that the very largeness of $M_{Pl}$ 
or the weakness of physical gravity could be due to hidden extra
dimensions, with the underlying (higher-dimensional) quantum gravity
scale being much lower and the underlying gravity therefore being
stronger. For a simple example of $d$ extra dimensions compactified on
circles of equal radius $R$, it is easily shown that:
$$
M^* = \frac{M_{Pl}}{(R M_{Pl})^{d/(d+2)}}
$$
Thus any value of $R M_{Pl} \gg 1$ will make $M^* \ll
M_{Pl}$. 

In some versions of this scenario, the Standard Model matter fields
propagate on a $3+1$-dimensional sub-manifold of the full spacetime
called a ``brane'', while gravity propagates in the entire ``bulk''
spacetime. Now for a compactification on a single circular dimension,
every field propagating in the bulk will display a characteristic
signature in the 3+1 dimensional world in the form of an infinite
tower of states spaced equally in mass in units of $R^{-1}$. For
addition circular dimensions or more general internal spaces, there
are similar characteristic regularities in the spectrum. These
graviton ``resonances'' are predicted by these models and should be
seen experimentally.

In the class of such models proposed by Arkani-Hamed, Dimopoulos and
Dvali (ADD), $M^*$ is taken to be of the order of a TeV. This requires
an $R \sim 10^{(32/d - 19)}$ metres, which amazingly satisfies all the
observational constraints for $d > 1$. For instance, $d=2, 3$ and 7
respectively imply $R \sim 1$ mm, 1 nm and 1 fm respectively. A
deviation from the inverse square distance dependence of Newton's
gravitational law is predicted at distances less than or close to
$R$. For 2 extra dimensions, torsion balance experiments suggest
that $M^*$ will lie above the reach of the LHC.

Now if $M^*$ is $\cal O$ (TeV), the underlying higher-dimensional
quantum gravity acts as a TeV-scale cutoff on the Higgs mass, in
contrast to the Planck scale cutoff in a conventional 4-d gravity
theory. Thus the naturalness (or hierarchy) problem associated to the
Higgs mass in the Standard Model is solved. As a bonus, one has the
exciting prospect of signals from these extra dimensions at TeV
energies, for example at the LHC. Specifically, with fairly tiny values of
$R^{-1}$ (e.g. $R^{-1} \sim 10^{-3}$ eV for $R
\sim 0.1$ mm), a whole tower of gravitonlike states, very closely
spaced in mass, will be produced there without being directly
detectable. However, an incoherent sum of these will contribute to
observable events with missing energy and momentum. 

Such events are predicted, for instance, with only one accompanying
hadronic jet in the final state: a configuration more characteristic
of the present scenario than of other beyond-Standard-Model scenarios. The
total cross sections for such ``monojet plus missing energy" events
have been calculated in these models and have been found to be
measurably large at LHC energies . There are additional signals
involving the exchange of virtual graviton resonances. Another
exciting possibility [53] in such models with a TeV scale quantum
gravity is that of producing mini black holes at the LHC at a huge
rate. These will almost instantaneously ($\sim 10^{-26}s$) evaporate
via Hawking radiation leading to an enormous number of photons as well
as other Standard Model particles in flavor-blind final states. The
current literature is rich with phenomenological discussions [54] of
the possible occurrence and detection of such events.

\subsubsection*{(ii) Warped Extra Dimensions}

The warped extra dimension idea, due originally to Randall and Sundrum
(RS), has spawned a vast collection of models. The compact fifth
dimension has a radius $R$ and can be parametrised by an angle $\phi$,
with $0<\phi<2\pi$ (thus the physical coordinate in this direction is
$R\phi$). The metric of the total spacetime is: $$ ds^2 =
e^{-2kR\phi}\eta_{\mu\nu}dx^\mu dx^\nu - R^2(d\phi)^2 $$ where
$\eta_{\mu\nu}$ is the standard Minkowski metric $diag (1,-1,-1,-1)$
and the exponential warp factor depends on a constant $k \sim M_{Pl}$. This metric corresponds
to a solution of Einstein's equations in five dimensions with a
specific negative value of the cosmological constant. This makes it an
anti-deSitter (AdS) spacetime. 

Additionally, two $3+1$-dimensional branes located at fixed values of
the transverse coordinate $\phi$ are postulated: one is called the ``Planck''
or ``ultraviolet'' brane and lies at $\phi =0$, while  the other is
called the ``infrared'' brane and is placed at $\phi =
\pi$. In the simplest models, all Standard Model fields are located on
the infrared brane and only gravitons propagate in the bulk. Both
branes extend infinitely in the usual three spatial dimensions, but
are thin enough in the warped direction so that their profiles can be
well-approximated by delta functions in the energy regimes of
interest. $M_{Pl}$-size operators on the ultraviolet brane lead to
laboratory energy effects on the Standard Model brane with a typical
scale of $\Lambda_\pi \equiv M_{Pl} e^{-kR\pi}$. This can now
broadly match with $M_W$ with the choice $kR \sim 12$. Thus, at the
cost of slightly fine-tuning the value of $kR$,
the hierarchy problem is taken care of.

The stability of the separation of the two (3+1) dimensional 
branes, or in other words the value $kR\sim 12$, has been investigated. This 
can be ensured by the presence of a spin-zero field in the bulk 
which needs to have a suitable potential. Characteristic experimental 
signatures [55] of RS models come from the spin-two Kaluza-Klein graviton 
resonances which should show up with mass spacings much larger than in the ADD 
case. In fact, the spacings here are in the range of hundreds of GeV. Moreover, 
their couplings to ordinary particles are of nearly electroweak strength since 
their propagator masses are red-shifted on the infrared brane. 

RS gravitons resonances can be produced from quark-antiquark or
gluon-gluon annihilation at the LHC, and in $e^+e^-$ annihiliation at
the proposed ILC. There are also specific signals [56] for the radion.

\section{String theory: beyond quantum fields}

\subsection{Physical origins of string theory}

The beginnings of the string idea can be traced to properties of the
strong interactions and the quark model that we have already
described. Experiments in the late 1960's indicated that hadrons are
indeed made up of quarks but these are permanently confined, at least
at zero temperature and in a vacuum environment. For mesons (such as
the pion), the constituents are a quark-antiquark pair. Confinement
could quite naturally be modelled by supposing that the quark and
antiquark are joined by a ``string'' with a constant tension. On
pulling the quark apart from its antiquark, the energy would grow
linearly, much like the energy in a rubber band, and the particles
would never separate. Since we now believe that the gauge field theory
of Quantum Chromodynamics (QCD) correctly describes the strong
interactions, this string would not be a truly independent object, but
could be created dynamically as a narrowly collimated tube of colour flux.

\begin{figure}[ht]
\begin{center}
\includegraphics[height=4cm]{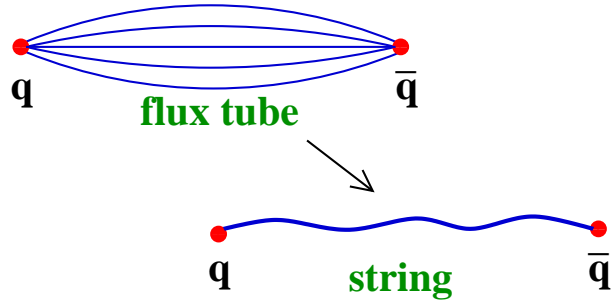}
\end{center}
\caption{Replacing the flux tube by a fundamental string.}
\label{fluxstring}
\end{figure}

A natural quantum-theoretic variation on the classical string picture
is that when the energy in the flux tube is large enough, it can break
by acquiring a quark-antiquark pair from the vacuum. The end result
is then a pair of mesons, one containing the original quark bound to
a new antiquark from the vacuum, the other containing the original
antiquark now bound to a new quark from the vacuum. This is consistent
with the experimental fact that when mesons (and more generally all
hadrons) are subjected to high-energy collisions, they fragment into
other hadrons whose quark content typically contains extra
quark-antiquark pairs relative to the initial state.

Early enunciations of this principle were due to Nambu and Susskind
following upon work of Veneziano [57]. They proposed in essence that
the flux tube be treated as a ``fundamental string''. Such a string
would be open, having two endpoints at which are attached a quark and
an antiquark. In this picture, confinement would be built in by
assigning the string a fixed tension $\tau_s$ with dimensions of mass
per unit length. In units where $\hbar=c=1$, this is equivalent to
inverse length squared. For historical reasons the string tension is
parametrised in terms of a quantity called $\alpha'$ with dimensions
of $({\rm length})^2$: $$
\tau_s = \frac{1}{2\pi\alpha'}
$$
The splitting of an open string into two parts, a natural interaction
to allow in such a theory, would inherently include the process of
popping a quark-antiquark pair out of the vacuum, since each open
string in the final state would have its own quark-antiquark pair at
the ends.

It was felt that the string description of hadrons would be more
useful in describing certain aspects of the strong interactions, but
possibly less useful to describe other aspects. In particular
perturbative QCD, which is well understood using quantum field theory
(as it involves familiar Feynman diagram techniques applied to study the
interactions of quarks and gluons), would not necessarily be easy to
understand in the string picture. However, phenomena related to
confinement would be easier to study, since confinement itself was
built in to the theory. In this sense it was implicit from the outset
that, if successful, string theory would provide a description of strong
interactions complementary to that provided by QCD. 

\subsection{Quantisation of free open strings: Regge behaviour}

These physical ideas led a number of researchers to take up the
quantisation of a relativistic fundamental string, propagating in flat
spacetime, as a theoretical challenge. Formulating a classical action
for a fundamental relativistic string is rather straightforward if one
uses analogies with relativistic point particles. In the latter case,
the natural invariant that plays the role of the classical action is
the invariant length of the particle's world-line.  This action is
invariant under two distinct symmetries: Lorentz transformations in
spacetime as well as arbitrary reparametrisations of the intrinsic
time parameter.

For a string one has a two-dimensional ``world-sheet'' rather than a
world-line, with the two dimensions corresponding to the spatial
location along the string and the intrinsic time parameter. The
natural geometric invariant is now the {\it area} of this sheet. Again
this is Lorentz-invariant in spacetime, and also invariant under
arbitrary reparametrisations of the two worldsheet parameters
together.  The latter is a large and interesting group (less trivial
than the reparametrisations of a single time parameter which arise
in point-particle theory).  To a large extent this local symmetry
group determines the properties of fundamental strings.  

For open strings, one needs to supplement the ``area law'' action with
boundary conditions for the end-points of the strings. The most
natural boundary condition (and the only one that preserves Lorentz
invariance) is the Neumann (N) boundary condition stating that the
end-points are free to move about in spacetime, but that no momentum
flows out of the end of the string. The other possible boundary
condition is Dirichlet (D), where the ends of the string are fixed to
a particular spacetime location. Each of the two endpoints of a string
can have independent boundary conditions, and also the boundary
conditions can be independently chosen as N or D for each direction of
space. D boundary conditions violate translation invariance and were
therefore largely ignored for many years, but they have turned out
crucial in understanding string theory dynamics, as we will see below.

Along with open strings, it is natural to quantise strings that are
closed on themselves with no endpoints. Indeed, since open strings are
allowed to interact by merging their endpoints, one expects a single
open string to be able to turn into a closed string after such an
interaction. Therefore theories involving open strings should always
have a closed string sector. In the string theory of strong
interactions, closed strings would correspond to states not involving
any quarks. From the microscopic perspective such states are
excitations purely of gluons, the force-carriers of the strong force, and are
therefore known as ``glueballs''. 

The quantisation procedure [58] is similar for open and closed strings, but
the latter case requires periodic boundary conditions in the parameter
that labels points on the string. For both open and closed strings, a
Hamiltonian treatment can be carried out by suitably gauge-fixing the
local worldsheet symmetries. Then the string reduces essentially to a
set of harmonic-oscillator modes for its transverse fluctuations. The
modes are labelled by an integer and correspond to ``standing waves''
on the open string of arbitrary integral wave number. Excited states
of the string are obtained by exciting each of these oscillator modes
an arbitrary number of times. In this way one finds an infinite
collection of states, each of which has a definite mass and angular
momentum. The spacing of the states goes like: $$ ({\rm mass})^2
~\sim~ \frac{n}{\alpha'} $$ for all possible integers $n$.

One of the first properties to emerge by quantising the open-string
world-sheet Hamiltonian (and which was already guessed at from the
physical picture of mesons as open strings), was ``Regge
behaviour''. This is the fact that string states lie on linear
trajectories when their angular momentum is plotted against the square
of their mass. This is easy to see qualitatively. The leading Regge
trajectory arises for states that are excitations of the lowest
standing wave. If we excite this mode $n$ times, we reach a state of
$({\rm mass})^2\sim n/\alpha'$ as above, but also (because each
excitation is a spacetime vector and therefore carries unit spin) the
state has angular momentum $J=n$. As a result we find: $$ ({\rm
mass})^2 ~\sim~ \frac{J}{\alpha'} $$ which is a linear Regge
trajectory with slope $1/\alpha'$.

Regge behaviour is observed experimentally. Large numbers of hadronic
resonances are found which, on a plot of $({\rm mass})^2$ versus spin,
lie on straight lines. 

\begin{figure}[h]
\begin{center}
\includegraphics[height=8cm]{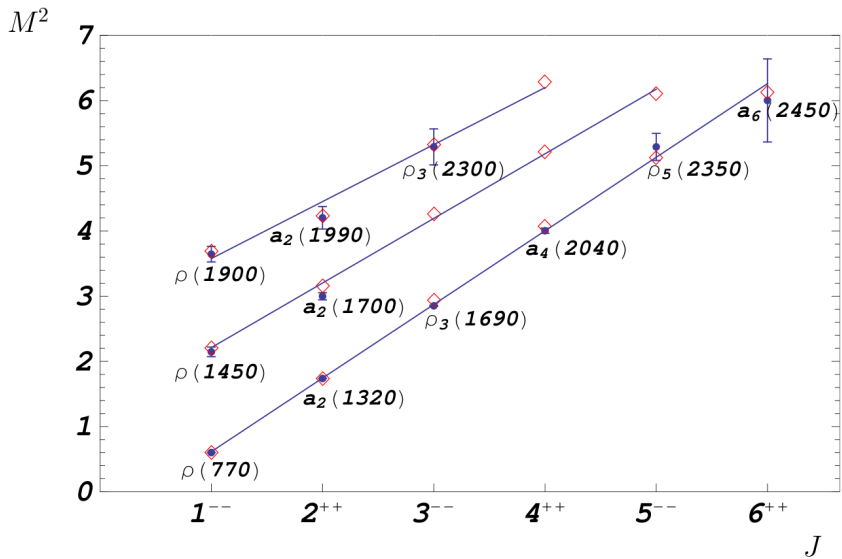}
\end{center}
\vspace*{-8mm}
\caption{Regge behaviour for a class of 
mesons. {\it (From D. Ebert et al, arXiv:0903.5183.)}}
\label{regge}
\end{figure}

This is a striking, though rather qualitative, experimental success
for the string description of hadrons. The proposal has proved very
hard to implement concretely, and to date this remains an unfinished
project, though remarkable progress has made following the AdS/CFT
correspondence, which we discuss in a later subsection.

\subsection{Critical dimension, tachyons, interactions} 

One experimental feature of Regge behaviour is that the straight line
trajectories have an ``intercept'', that is, they cut the
angular-momentum axis at a positive value. Thus a massless particle
will have nonzero angular momentum. Moreover if the line is produced
further, it reaches vanishing angular momentum at a point of negative
$({\rm mass})^2$, or imaginary mass. Such a point represents an
instability of the perturbative vacuum of the theory, the
corresponding particle being labelled a {\em tachyon} since it travels
faster than light. If an intercept $a$ is present in string theory
then there will be a tachyon of $({\rm mass})^2=-a/\alpha'$.

\begin{figure}[h]
\begin{center}
\includegraphics[height=5.5cm]{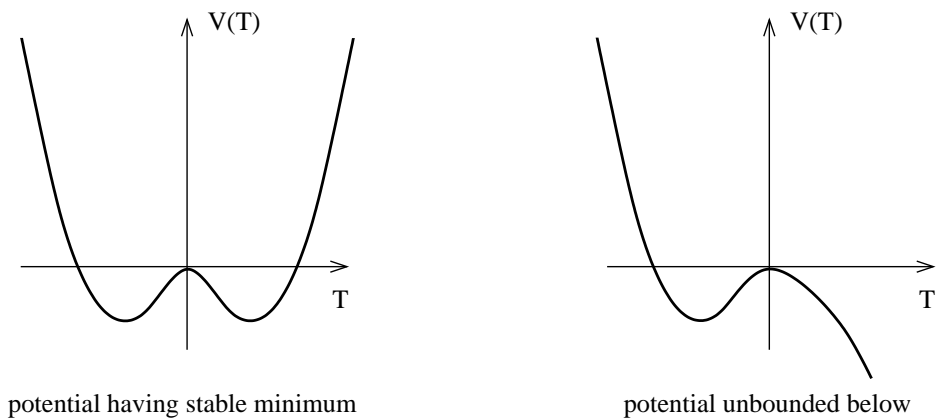}
\end{center}
\vspace*{-2mm}
\caption{Two possible types of tachyon potentials, one having a stable
minimum and the other being unbounded below.}
\label{tachyon}
\end{figure}

The negative $({\rm mass})^2$ means the field $T$ associated to this
particle has a potential with a local maximum at the origin. If the
potential has a stable minimum at some other value, the tachyon field
will ``roll down'' to this minimum and the theory will be well-defined
and tachyon-free at the end. However if the potential is bottomless
then the instability is serious and the theory is deemed to be
inconsistent (see Fig.\ref{tachyon}).

Careful quantisation of the string world-sheet theory reveals,
purely on theoretical grounds, that an intercept must be present. The
world-sheet theory has potential problems with either unitarity or
Lorentz invariance, depending on the quantisation scheme, and in order
to preserve both of these physically essential properties one is
obliged to fix two parameters: the dimension $D$ of the spacetime in
which the string propagates, and the intercept $a$. A variety of
different quantisation schemes all yield a common answer: for open
strings, $D=26$ and $a=1$. Things are not so different for closed
strings: one again finds $D=26$ and this time the intercept is $a=2$.

Thus, relativistic ``bosonic'' strings\footnote{They are called
``bosonic'' because their coordinates are ordinary commuting
coordinates, to differentiate them from ``fermionic'' strings with
extra anti-commuting coordinates that we will shortly discuss.} are
consistent only in 26 spacetime dimensions and even there, both open
and closed strings possess tachyonic particles in their
spectrum. Despite these problems, the theory possesses some remarkable
features.  Among the infinitely many excited states of the string, the
massless states are of particular interest. Open strings are found to
contain in their spectrum a massless particle of spin 1 (in units of
the Planck constant), while closed strings contain a massless particle
of spin 2. Now it has long been known in relativistic field theory
that consistent theories involving spin 1 and spin 2 massless
particles are possible only if, in the former case, there is gauge
invariance, and in the latter case, there is general coordinate
invariance. Moreover, in the latter case the spin-2 particle can be
identified with the graviton, the force carrier of gravity. Thus
either string theory contains within itself the information about
these two local invariances, or else it will fail to be consistent. To
decide between these alternatives, it is essential to go beyond free
strings and understand string interactions.

For this purpose, it is useful to recast the world-sheet theory in the
language of dynamically fluctuating random surfaces. In a quantum
treatment, one describes the string in terms of a path integral over
random surfaces. These surfaces can have different topologies. In
fact, ``handles'' in the worldsheet turn out to correspond to
higher-loop diagrams in field theory language and punctures of the
worldsheet correspond to the insertion of asymptotic external
states. Thus the worldsheet formalism is able to compute for us not
only the spectrum of a free string but also, almost as a bonus, the
scattering amplitudes for arbitrary numbers of external string states.

Along with the massless spin-2 particle described above, closed
strings also give rise to a scalar particle called the ``dilaton''
$\Phi$ with a special property. Its constant vacuum expectation value
acts as the coupling constant $g_s$ in string theory:
$$ e^{\langle \Phi\rangle} = g_s $$ This interpretation follows from
the fact that the power of $g_s$ accompanying the amplitudes is
proportional to the genus of the surface. Thus a precise
procedure exists (in principle) to compute string scattering
amplitudes to any order in perturbation theory in $g_s$.

These amplitudes in turn can be converted into interaction terms in an
effective Lagrangian. The result depends on the background in which
the string is propagating, including the metric of spacetime.  Indeed,
given the background, the interactions are completely determined. In
this sense, string interactions are fixed internally by the
theory, very unlike the case in usual quantum field theory where
interactions are inserted in the Lagrangian by hand.

In quantum field theory, the so-called ``loop diagrams'',
corresponding to higher-order contributions to perturbative scattering
amplitudes, are generically ultraviolet divergent. This led many
decades ago to the introduction of the renormalisation programme,
specifically for the theory of quantum electrodynamics. Today this
programme is now known to work quite generically for gauge theories
including non-Abelian ones, its success being intimately tied to gauge
invariance [15]. But it does not work for gravity, which is therefore
believed to be a non-renormalisable theory. By contrast, in string
theory studies of the loop diagrams clearly indicate that
ultraviolet divergences are generically absent. While this is a
solid formal result, there is also a simple physical reason for it,
namely the spatial extent of the string $\sim\ell_s$, which provides
an intrinsic ultraviolet cutoff on all short-distance processes.

Computing the interaction terms (in a flat spacetime background) for
the spin-1 and spin-2 massless particles of open and closed strings,
one finds the astonishing result that the self-interactions of the
former turn out to possess (Abelian) gauge symmetry, while the latter
possess general coordinate invariance. In fact the scattering
amplitudes of the spin-2 massless state of the closed bosonic string
precisely reproduce the expansion, in powers of the metric fluctuation
about flat space, of the famous Einstein-Hilbert Lagrangian for
gravitation! The result is particularly noteworthy for the fact that
no information about gravitation or general coordinate invariance was
put in to the theory at the outset.

The above results quickly made closed string theory a natural
candidate for a quantum theory of gravity. This marked a departure
from the original goal of using strings to describe confinement in
strong interactions. In the process the string tension had to be
re-calibrated so that 
$$ 
(\ell_s)^{-1}\sim 10^{19}~ \hbox{Gev} 
$$ 
as against the strong-interaction value of this quantity which was in
the range of 10-100 Gev. One could now have a theory that naturally
contained gravitation and was ultraviolet finite, one of the most
elusive goals in field theory since the days of Einstein.

Despite convincing evidence for ultraviolet finiteness, the actual
computation of loop amplitudes in bosonic string theory encountered a
stumbling block because of the tachyonic instability referred to
earlier. The presence of the tachyon introduced new types of
divergences into loop diagrams. At this stage it became necessary to
understand whether the tachyon potential has a stable minimum or
not. This is a technically hard question, but it now seems most likely
that the closed-string tachyon indeed has a bottomless potential, as
on the right side of Fig.\ref{tachyon}, which would mean that the
bosonic string theory is inconsistent even in 26 dimensions.

Fortunately the positive features, notably gauge and gravity fields,
of string theory have survived in a modified version called
``superstring theory'', that seems to be free of tachyons and other
inconsistencies.  Hence today ``string theory'' usually refers to the
superstring variant, while the bosonic string has fallen into disuse
except as a pedagogical arena to understand some of the basic notions
of string theory.

\subsection{Superstrings}

The addition of new fermionic degrees of freedom to the string in
addition to the usual (bosonic) spacetime coordinates turned out to be
a key idea that remedied most of the defects of the theory while
preserving its positive features. With these new degrees of freedom,
it is possible to implement supersymmetry on the string
worldsheet\footnote{Indeed, supersymmetry was first discovered in this
context [59].}. This in turn changes the computations of the critical
dimension and intercept in a crucial way, the new values being $D=10$
for the critical dimension and $a=0$ for the intercept. In the
bargain, one ends up with a theory which is supersymmetric in
spacetime and is called ``superstring theory''\footnote{It is to be
noted that supersymmetry on the worldsheet (which is an auxiliary
space) is not physically observable and differs from the potentially
observable supersymmetry in spacetime which causes every bosonic
excitation of the string to possess a fermionic partner.}.

Because of the vanishing intercept, superstring theories have no
tachyon. Moreover the critical dimension is lowered to a point where
one makes contact with previous studies higher-dimensional
gravitational theories in a Kaluza-Klein framework [60]. We defer
discussion of this to a later subsection.

At tree level (lowest order in perturbation theory), adapting the
technique of summing over random surfaces to superstring theories one
again finds spin-1 and spin-2 particles whose self-interactions
indicate the presence of gauge invariance and gravity
respectively. There are additional ``superpartner'' particles as
required by supersymmetry. Indeed, fermionic particles now make their
appearance for in string theory for the first time, for they were
completely absent in the bosonic string. Now that there is no tachyon
one can compute one-loop corrections, and here it is quite
conclusively seen that the theory produces completely finite loop
amplitudes for gravitons in flat spacetime. Everything that has been
learned about superstring theory since then has supported the proposal
that it is an ultraviolet finite and consistent theory of quantum
gravity.

Careful investigation revealed that there are precisely five distinct
superstring theories, all consistent only in 10 dimensions. Two of
these have the maximal allowed supersymmetry in 10 dimensions, and are
known as ${\cal N}=2$ (corresponding to 32 independent supersymmetry
charges). They are labelled type IIA/IIB superstrings.  The remaining
three string theories have half this amount, or ${\cal N}=1$
supersymmetry, or 16 supercharges.

For each of the two maximally supersymmetric cases, a unique classical
field theory possessing all the desired symmetries (``type IIA/IIB
supergravity'') is known in 10 dimensions. When we consider strings in
the $\alpha'\to 0$ limit of slowly varying backgrounds relative to the
string scale, string theory should reduce to ordinary field theory. In
this limit, symmetry considerations alone tell us that the field
theory must be type IIA/IIB supergravity. Similar considerations hold
for the ${\cal N}=1$ supersymmetric or ``type I'' strings. Here,
however, the supersymmetries allow for coupling of supergravity to
gauge fields and there is therefore a gauge group to be chosen. This
choice appears arbitrary (as it usually is in four-dimensional field
theories) until one takes into account subtle quantum anomalies that
occur due to the chiral or parity-violating nature of the theory. In a
landmark paper in 1984, Green and Schwarz [61] discovered that the only
allowed gauge groups are $SO(32)$ and $E_8\times E_8$. The three
${\cal N}=1$ supersymmetric string theories correspond to these two
gauge groups, with two string theories having $SO(32)$ and one having
$E_8\times E_8$ gauge group.  The particularly natural way in which
the latter can reproduce many features of the real world upon
compactification to four dimensions caused enormous excitement in 1984
and led to an explosion of papers on the subject [62].

The fact that string theory reduces to field theory as $\alpha'\to 0$,
and more specifically that superstring theories reduce to very
well-known supergravity theories in the same limit, has an important
impact on the way we think about the subject. First of all, it means
that relatively esoteric techniques are not necessarily required to
understand string dynamics at low energies, since effective field
theory suffices for this purpose and our intuition about it has been
honed by over half a century of experience. Second, the domain where
stringy effects become important, namely at very high (Planckian)
energies, is one in which field theory simply does not work, as
evidenced by the fact that direct quantisation of Einstein's action is
inconsistent at loop level, and it is here that ``stringy'' features
show up and ``stringy'' techniques are indispensable.  This may
suggest that experiments at present-day energies will never be able to
test string theory, and there is some truth in this belief within the
most obvious scenenarios (it would of course be equally true for any
other theory that we might propose to describe quantum gravity at the
Planck scale). However there is some optimism on a variety of fronts,
an important one being in cosmology where Planck-scale energies could
actually be probed by high-precision experiments. There are also
unconventional proposals suggesting that the 10-dimensional Planck
scale is much lower than commonly assumed and therefore
higher-dimensional behaviour might become accessible at accelerators.

In any case, rather than being viewed as a departure from
familiar physics, string theory should be viewed as a natural
modification and extension of quantum field theory into a domain where
the latter is inapplicable. This is similar in some ways to the fact
that quantum mechanics allows us to modify and extend classical
mechanics into the short-distance domain, where the latter simply
fails to apply. The analogy is strengthened by comparing the role of
the deformation parameters $\hbar$ and $\alpha'$:
\begin{eqnarray}
\hbox{Quantum Mechanics }&\stackrel{\hbar\,\to\, 0}{\longrightarrow}& 
\hbox{ Classical Mechanics}\nn\\
\hbox{String Theory }&\stackrel{\alpha'\,\to\,0}{\longrightarrow}& 
\hbox{ Quantum Field Theory}\nn
\end{eqnarray}

\section{String compactification}

\subsection{Simple compactifications}

Having obtained a consistent quantum theory of gravity and gauge
interactions in 10 dimensions via superstrings, one has already
achieved a measure of success over previous attempts to quantise
gravity. Loop amplitudes for graviton-graviton scattering are
calculable in 10D superstring theory for the first time.  All the
indications are that superstring theory is unitary and passes all
theoretical tests for a sensible theory. However one would like to
extend this success by using superstrings to describe a physical
system in 4 spacetime dimensions that has the structure of the
Standard Model, augmented by gravity. 

Compactifying superstring theory {\em per se} is quite natural. The
consistency requirement discussed above only fixes the number of
spatial dimensions to be 9, but does not require them to all be
infinite in extent. Therefore it is possible to choose some dimensions
(say 6 of them) to be compact. The idea that space has ``hidden''
extra dimensions that are compact and therefore unobservable at low
energies, and that this may in some way unify gauge and gravitational
interactions, is nearly a century old and dates from work of
Th. Kaluza and O. Klein. String theory has brought about a
resurrection of this idea. And importantly, it has for the first time
rendered it viable in the context of quantum and not merely classical
theory.

A convenient and relatively simple starting point for superstring
compactification is to consider the corresponding low-energy
supergravity theories in 10 dimensions. Simple but unrealistic
compactifications are easy to find, by choosing an internal manifold
that is a product of 6 circular dimensions. These give rise to
4-dimensional theories with experimentally undesirable features such
as enhanced supersymmetry and parity conservation, and more generally
an inappropriate particle content compared with the real
world. Nevertheless even these unrealistic compactifications teach us
important lessons. As an example, within these one can compute
graviton-graviton scattering amplitudes -- including loop amplitudes
-- in a four-dimensional setting, demonstrating that the theoretical
successes of superstrings are not limited to 10 dimensions.

\begin{figure}
\begin{center}
\includegraphics[height=2.5cm]{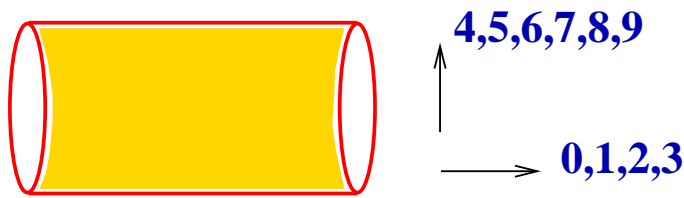}
\end{center}
\caption{Schematic depiction of six compact and four non-compact dimensions.}
\label{compact}
\end{figure}

Before discussing some details of string compactifications, let us
also note an important feature of 10-dimensional superstrings that
survives in their compactifications to 4 dimensions. This is the fact
that gravity and gauge interactions have a common origin. Experiment
tells us that all fundamental forces in nature are either of gauge or
gravitational type. String theory has two types of strings, closed and
open. There is a perfect match: closed strings are the origin of
gravity and open strings the origin of gauge forces\footnote{This is
strictly correct in the type II string theories. In the heterotic [63]
theories all forces arise from closed strings.}. It is no surprise
that compactified string theory has come to be seen as the most
promising candidate not only for quantum gravity but also for the
unification of all forces in nature.

It should also be noted that a compactified string theory is not a
different theory from the original one in 10-dimensional flat
spacetime, rather both are to be considered different vacua of the
same underlying theory. 

\subsection{Early compactifications and moduli}

Modern variants of the Kaluza-Klein idea, in vogue in the early
1980's, had already zeroed in on 6 extra dimensions for a variety of
phenemenologically inspired reasons, not least the fact of parity
violation in the real world. These developments were put to use in a
landmark paper by Candelas, Horowitz, Strominger and Witten [64] in
1984, who produced a family of compactifications whose starting point
was the ``$E_8\times E_8$ heterotic string'' in 10 dimensions, and for
which the 6-dimensional compact manifold was one of a large class of
``Calabi-Yau'' manifolds that mathematicians had previously
studied. The compact manifold was a classical solution of the
supergravity equations of motion. These compactifications led to a
class of four-dimensional field theories with the following (at the
time) desirable properties: Minkowski spacetime with vanishing
cosmological constant, parity violation, a realistic gauge group and
replicating fermion generations. Moroever the compactified theories
had ${\cal N}=1$ supersymmetry in 4 dimensions, a phenomenologically
desirable extension of the Standard Model\footnote{The presence of
${\cal N}=1$ supersymmetry at the TeV scale will be tested by the
Large Hadron Collider.}.

Soon thereafter, it became clear that the conditions for a consistent
compactification could be understood directly in string theory as
consequences of conformal invariance on the string worldsheet. Strings
propagating in curved spacetimes were described by a nontrivial
worldsheet theory [65]. Quantum effects typically violated
conformal invariance unless certain conditions were satisfied
[66] and these conditions could be reinterpreted as
classical equations of motion for the string\footnote{The role of conformal
invariance in replacing classical equations of motion has led to some
novel ``compactifications'' in which the internal
space is not a geometric space at all, see for example
[67] and references therein.}.

The compactifications studied in the initial works of this period
had several limitations and were therefore not yet realistic. A
key limitation was that for every deformation mode (the mathematical
terminology is ``modulus'') of the internal Calabi-Yau manifold, a scalar
particle with a vanishing potential, and therefore in particular
zero mass, would necessarily exist in the compactified theory, in
flagrant contradiction with experiment. Most compactifications indeed
gave rise to hundreds of such ``moduli'' particles. It was expected
that some mechanism would ``lift'' the moduli by generating a
potential stabilise them, but work on this question proceeded somewhat
slowly, partly because the new theoretical tools needed for essential
progress had not yet been discovered.

With the advent of D-branes in the mid 1990's and the subsequent
landmark discovery of the gauge/gravity correspondence (both are
discussed in the following section), string compactification received
a new impetus. Its goals could now be addressed using novel theoretical
tools. It also became clear on the experimental side that there is a
``dark energy'' most plausibly described by assuming that gravitation
includes a small positive cosmological constant. Thus instead of
seeking a compactification to Minkowski spacetime, the goal
was revised to finding string compactifications to asymptotically deSitter
spacetime in four dimensions.

The culmination of many attempts in this direction came with a classic
work by Kachru, Kallosh, Linde and Trivedi [68] in which recent
theoretical developments in string theory were successfully
incorporated into a class of compactified models. Their models start
with ``warped'' internal spacetimes with fluxes turned on, locally
resembling Anti-deSitter spacetime as in the gauge/gravity
correspondence (see following section). Besides the fluxes, which
stabilise many of the moduli, nonperturbative effects are invoked to
stabilise the remaining moduli. Inserting anti-branes (the D-brane
analogues of antiparticles) into the warped geometry has the effect of
breaking supersymmetry and lifting the cosmological constant to a
positive value. The resulting models were grossly similar to the
standard model and had no massless particles.

It soon became clear that string theory admits an enormous number
(sometimes estimated as $10^{500}$ [69]) of consistent vacua
fulfilling basic physical requirements. This led to the problem that
there could be an enormous number of vacua describing worlds
arbitrarily close to the real world, hence it would be virtually
impossible to find ``the right one''. Moreover it then becomes hard to
see what principle selects one out of such a large multiplicity of
vacua. Some scientists support the ``anthropic principle'' according
to which one favours the kind of vacua that admit the possibility of
human life, but others find this approach unscientific (a lucid
exposition of the principle for the layman can be found in
Ref.[70]). It is fair to say that this issue is not conclusively
settled at present.

\section{Duality, D-branes and the gauge/gravity correspondence}

\subsection{Solitons, duality and D-branes}

In the early 1990's it was found that classical string theory has a
spectrum of charged stable solitons that are typically extended in one
or more spatial directions. Since these were generalisations of
membranes to higher dimensions, with similar properties to black
holes, they were dubbed ``black branes''[71]. They were discovered as
classical solutions of the effective low-energy supergravity theory,
possessing both flux (because of their charge) and a gravitational
field (because of their tension). Their existence was of interest, if
not entirely unprecedented, because they possessed some elegant
theoretical properties including the preservation of upto half the
supersymmetries of the underlying theory. Additionally they filled an
important gap in type II superstring theories: the perturbative string
spectrum of these theories included some massless tensor gauge fields
(analogous to photons, but with additional vector labels) that are for
technical reasons known as ``Ramond-Ramond'' fields. But surprisingly,
there did not seem to be any states in the theory that were charged
under these gauge fields. This anomalous situation was resolved when
some of the extended solitons turned out to carry Ramond-Ramond
charges, one consequence of which was to render them stable.

\begin{figure}
\begin{center}
\includegraphics[height=4.5cm]{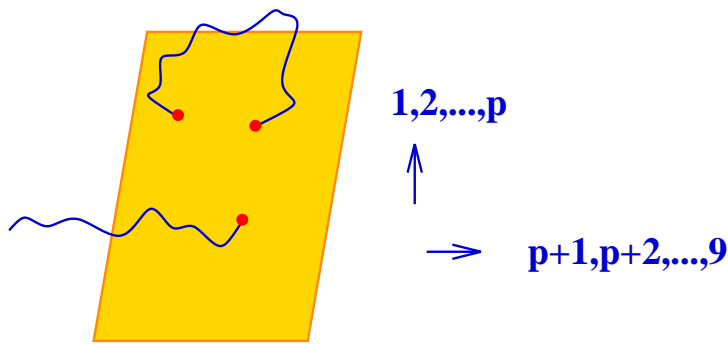}
\end{center}
\caption{A planar D$p$-brane, showing allowed open strings and spatial directions.}
\label{directions}
\end{figure}

Solitons, such as vortices or monopoles, are nonperturbative objects
in field theory. The techniques available to study them are typically
somewhat limited, as even lowest order perturbation theory is
relatively difficult to carry out in nontrivial backgrounds. Hence
this discovery attracted only moderate interest for some years, until
a path-breaking observation was made by Polchinski in 1995 [72]. He
carefully studied open strings whose end-points have Dirichlet
boundary conditions along some subset of the 9 spatial directions and
Neumann boundary conditions along the remaining directions. Such
boundary conditions cause the string end-points to be ``anchored'' on
a hypersurface of dimensionality equal to the number of Neumann
boundary conditions. This defines a ``wall'' of any desired
codimension in space, as depicted in Fig. \ref{directions}. Polchinski
provided a convincing argument that such walls were actually dynamical
objects, and he used open-string dynamics to compute their tension. He
also showed that some of these ``Dirichlet branes'' or ``D-branes'' as
they came to be known, carried Ramond-Ramond charges. It was then
natural to identify them with the classical solitons carrying RR
charge that had already been discovered.

In this sense the objects were not new, but the fact that their
dynamics could be completely described by open strings was novel and
would ultimately come to have an enormous impact on several branches
of physics. In particular, the origin of non-Abelian gauge symmetry in
open-string theories now became very natural, since a collection of
$N$ parallel D-branes supports $N^2$ distinct open strings (each
end-point can independently lie on one of the branes). These convert
the massless excitations of the string into a matrix-valued gauge
field. String interactions then automatically produce interactions of
Yang-Mills type.

A parallel development, also inspired by the initial discovery of
branes as classical solitons in string theory, was the growing
realisation that certain compactifications of type II superstring
theory possessed nonperturbative ``duality'' symmetries interchanging
strong and weak coupling [73]. The presence of such symmetries was, and
still is, generally impossible to ``derive'', as they exchange a
weakly coupled (and therefore accessible) theory with a strongly
coupled (and therefore ill-understood) theory. Nevertheless impressive
evidence could be accumulated in favour of the duality symmetries.  In
particular, they exchanged fundamental string states with solitonic
states, and therefore the discovery of solitons carrying the
appropriate charges made for a strong case in support of
duality. Similar dualities also appeared in highly supersymmetric
quantum field theories where they exchanged electric fundamental
particles with solitonic magnetic monopoles.  Supersymmetry was
invoked to guarantee that solitons, as also the fundamental string
theory/field theory states, existed both for weak and strong coupling,
so that they could be meaningfully compared.

Synthesising all the above developments, it finally emerged that the
five different string theories (type IIA, IIB, I, and the two
heterotic theories) were really different ``corners of parameter
space'' of a single common theory. This unification of all string theories
was deeply satisfying and reasserted the role of string theory as a
fundamental framework. Moreover, a sixth and very surprising corner of
parameter space was found that was not a string theory at all! It was
a theory involving membrane excitations and having a low-energy
description in terms of 11-dimensional supergravity. This theory,
dubbed ``M-theory'', is sometimes considered to be the master theory
underlying all string theories. The dynamics of membranes appears to
play a key role in this theory and has been the subject of significant
recent developments [74].

\subsection{Black hole entropy}

The discovery of strong-weak dualities and D-branes provided impetus
to an old programme to understand the nature and fundamental meaning
of black hole entropy. The famous Bekenstein-Hawking result stated
that black holes appeared to have an entropy proportional
to their horizon area (with a precisely known coefficient), and this
entropy obeyed the laws of thermodynamics. What was not clear until
the mid-1990's was whether this thermodynamic behaviour was (as 
in conventional statistical mechanics) simply the result of
averaging over an ensemble with microscopic degrees of freedom, or
pointed to some new physics altogether.

String theory being established as a consistent theory of quantum
gravity, it was a natural place to try and settle this question. In a
definitive work [75] it was shown that a class of black holes in
string theory can be described, for some range of parameters, as
D-brane bound states and that these can be counted accurately and give
an entropy in exact agreement with the Bekenstein-Hawking formula
including the numerical coefficient\footnote{A remarkably prescient
work in this direction, pre-dating the discovery of D-branes, was
published earlier by Sen [76].}. Following this dramatic result, a
number of generalisations were found. Additionally, decay rates for
black holes in the macroscopic picture (via Hawking radiation) and the
microscopic picture (via emission of quanta by branes) were compared
and found to agree [77]. These successes mean that string theory has
passed an important test required of any candidate theory of quantum
gravity, namely to shed new light on the famous paradoxes involving
black holes.

\subsection{The gauge/gravity correspondence}

A novel correspondence, that has come to dominate many areas of
theoretical physics over the last decade, was proposed in 1997 by
Maldacena [78]. The argument starts with the equivalence between black
branes (gravitating solitons) and D-branes (open-string
endpoints). Both sides of this equivalence admit a low-energy limit in
which the string length $\ell_s$ is scaled to zero keeping energies
fixed. For black branes, this limit keeps only the ``near-horizon''
region, which is the 10-dimensional spacetime $AdS_5\times S^5$. Here
the first factor represents five dimensional Anti-deSitter
spacetime\footnote{This is a highly symmetric spacetime with constant
negative curvature. A key feature of this spacetime is that it
gives rise to a negative cosmological constant.}, while the second factor is a
five-dimensional sphere. Applying the same low-energy to D-branes, it
is found that the surviving part of the action describes a maximally
supersymmetric Yang-Mills gauge theory in four dimensions.

Thus it was proposed that superstring theory in an $AdS_5\times S^5$
spacetime is exactly equivalent to maximally supersymmetric
(technically: ${\cal N}=4$ supersymmetric) Yang-Mills theory in
four-dimensional Minkowski spacetime. Even though it may sound
unlikely that a 10-dimensional string theory can be equivalent to a
4-dimensional quantum field theory, a huge body of evidence has by now
been amassed in favour of the correspondence, starting with a
comparison of symmetries, spectra and other basic properties. A key
role is played by conformal symmetry of the Yang-Mills theory, which
is realised as an isometry of anti-deSitter space.

The duality has two classes of applications: 

(i) Starting with the Yang-Mills theory, it offers a potentially
complete nonperturbative description of a string theory incorporating
quantum gravity. Thereby it promises to teach us fundamental things
about quantum gravity. 

(ii) Starting with the string theory, it offers the possibility of studying
non-Abelian gauge theories in a regime of parameter space where they
are hard to understand directly, and specifically opening a window to
the study of quark confinement.

In most practical applications to date, the string-theoretic aspects
of the AdS theory are not even used, as one works in the limit of
large AdS radius $R$ (and  weak string coupling) in which string theory
reduces to classical supergravity. The price one pays for this
simplification is that it applies only when the gauge theory has a
large number of colours $N\gg 1$ and the 'tHooft coupling $g^2 N$
is also large.

The physics of black holes in AdS spacetime plays an essential role in
learning about confining gauge theories. A key result in this area is
that if one continues the gauge theory to Euclidean signature and
introduces a temperature, then there are two distinct AdS-like
spacetimes dual to the gauge theory, one valid for low temperatures
(``thermal AdS spacetime'') and the other for high temperatures (a
black hole in AdS spacetime). Next, it can be argued that there is a
phase transition between the two spacetimes at some intermediate
temperature $T_c$. This phase transition corresponds to the formation
of black holes above $T_c$, which -- using the Bekenstein-Hawking
formula -- is seen to release a large amount of entropy $S\sim R^8$
where $R$ is the size scale of the AdS space. Translating the result
back to the gauge theory, one finds that this entropy is of order
$N^2$ where $N$ is the number of colours. This is the expected result
if $T_c$ is the deconfinement temperature above which free
quarks/gluons are liberated. It is extremely striking that black hole
formation and quark confinement, two of the most subtle and
fascinating phenomena in physics, are for the first time linked to
each other in the AdS/CFT correspondence [79].

Admittedly the super-Yang-Mills theory used in this discussion is
rather far from true QCD in its dynamics, but appears to fall in a
similar universality class when the temperature and other parameters
are chosen appropriately. Considerable work has gone into making more
sophisticated dual pairs where the gauge theory is closer to true
QCD. The holy grail, an analytic proof of confinement in true QCD,
still remains elusive at the time of writing this article.

A more recent application of the correspondence is carried out not in
Euclidean but directly in Minkowski signature, still at finite
temperature. In this case the appropriate dual is argued to be a black
3-brane in AdS spacetime. This object has a horizon whose dynamics can
be described using a previously known ``membrane paradigm'' for black
holes. On the field theory side, at finite temperature and long
wavelengths the natural description is in terms of hydrodynamics of a
plasma. Thus in this situation one uses the gravitational system to
learn about the properties of a gauge theory fluid, potentially
similar to the one created in ultra-relativistic high-energy
collisions. Gravity offers a powerful way to calculate transport
coefficients such as shear viscosity, for the fluid, and to derive
the nonlinear equations of boundary fluid dynamics [80]. As with the
earlier discussion on confinement, the duality here is best understood
for a fictitious fluid made up of the quanta of ${\cal N}=4$
supersymmetric gauge theory rather than true QCD.  However, the
remarkably useful predictions from this approach make it likely that
at least some of the results are applicable to the real world, and
that with more sophisticated models it might be possible to study true
quark-gluon plasma [81] in this way.

\section{Concluding remarks}

We have tried in this article to give a panoramic view of the main
theoretical developments which have taken place in high energy physics
since its inception more than half a century ago. During this period
experimenters have probed the energy range from a giga-electron-volt
to a tera-electron-volt, while theorists have come up with the
Standard Model which successfully explains most of the acquired
data. The Standard Model does, however, have inadequacies which have
led to new proposed theoretical schemes predicting characteristic
experimental signals at the TeV scale. There is also a linkage with
physics at very large distances of the order of giga light-years
through speculated weakly interacting massive particles forming
cosmological dark matter. 

There has been a paradigm shift from field theory to string theory,
which potentially provides a framework to explain quantum gravity and
to unify it with the other three interactions of the Standard
Model. In addition, string theory provides novel techniques to address
important dynamical questions about the strong interactions.

With higher tera-scale energies about to be probed by
the forthcoming Large Hadron Collider, the field of high energy
physics is vibrant with expectations for a new breakthrough.

\section*{Acknowledgements}

We are grateful to Gautam Bhattacharyya, Asit De, Rohini Godbole and
Rajaram Nityananda for their comments and suggestions. S. M. and
P. R. have been supported in part by a DST J. C. Bose and a DAE Raja
Ramanna Fellowship respectively.

\bigskip

{\LARGE \bf References}\footnote{The list of references given here is
indicative and far from complete, being intended primarily to give the
reader a few pointers to the literature rather than to properly credit
all original work.}

\begin{itemize}

\item[{[1]}] S. Raychaudhuri 2008, {\it A journey into 
inner space - the story of particle 
physics from its beginning to the eve of the LHC run}, 
Phys. News {\bf 38}, 7.

\item[{[2]}] S. Dodelson 2003, {\it Modern cosmology}, Academic Press, Amsterdam.

\item[{[3]}] {\it LHC: The Large Hadron Collider},
http://lhc.web.cern.ch/lhc/.\\ G. Kane and A. Pierce 2008 {\it
Perspectives on LHC physics}, World Scientific, Singapore

\item[{[4]}] C.N. Yang and R.L. Mills 1954,
Phys. Rev. {\bf 96}, 191.\\ 
R. Shaw 1955, Ph. D. dissertation, Cambridge.\\
Research on non-Abelian gauge theories
in India was initiated in:\\
G. Rajasekaran 1971, {\it Yang-Mills fields and 
theory of weak interactions}, 
lectures given at the Saha Institute of Nuclear Physics,
TIFR/TH/72-9.\\ 
T. Dass 1973,  {\it Gauge field theories} in {\it Advances in high energy 
physics}, Proceedings of the Summer School in Dalhousie, 1973, published by 
the Tata Institute of Fundamental Research.

\item[{[5]}] V. Singh and B. M. Udgaonkar 1963, Phys. Rev. {\bf 130}, 1177. 

\item[{[6]}] See e.g. S. M. Roy 1971, Phys. Lett. {\bf B 36}, 353. 

\item[{[7]}] See e.g. P. Roy 1974, Phys. Rev. {\bf D9}, 2631. 

\item[{[8]}] J. J. Kokkedee 1983, {\it The quark model}, Benjamin, New York. 

\item[{[9]}] M. E. Peskin and D. V. Schroeder 1995, {\it An introduction to quantum 
field theory}, Addison-Wesley, Redwood City. 

\item[{[10]}] P. Roy 1974, {\it Theory of lepton-hadron processes at 
high energies}, Clarendon, Oxford. 

\item[{[11]}] J. C. Pati and A. Salam 1973, Phys. Rev. {\bf D8},
1240.\\
G. Rajasekaran and P. Roy 1975, Pramana J. Phys. {\bf 5}, 303. 

\item[{[12]}] D. J. Gross 2005, Nobel lecture, Rev. Mod. Phys. {\bf
77}, 837.\\ 
F. Wilczek 2005, Nobel lecture, {\it ibid.}, 857.\\ H. D. Politzer 2005, Nobel lecture, 
{\it ibid.}, 851.  

\item[{[13]}] S. Banerjee 2004, Eur. Phys. Jour. {\bf C33}, S410. 

\item[{[14]}] S. L. Glashow 1980, Nobel lecture, Rev. Mod. Phys., {\bf 52}, 539.\\ S. 
Weinberg 1980, {\it ibid.}, 515.\\ A. Salam, 1980 {\it ibid.}, 525. 

\item[{[15]}] G. 't Hooft 2000, Nobel lecture, Rev. Mod. Phys., {\bf 72}, 332.\\ M. Veltman, 
{\it ibid.}, 341. 

\item[{[16]}] M. Kobayashi 2008, Nobel lecture,
http://nobelprize.org/nobel\_prizes/laureates/2008/kobayashi-lecture.html.\\
T. Maskawa 2008, {\it ibid.}/maskawa-lecture.html.

\item[{[17]}] G. Altarelli 2008, {\it New physics and the LHC}, 
arXiv:0805.1992 [hep-ph]. 

\item[{[18]}] V. Ravindran, J. Smith and W. L. van Neerven 2003, Nucl. Phys. {\bf B665}, 
325 

\item[{[19]}] G. Altarelli 1994, {\it The development of perturbative QCD}, World 
Scientific, Singapore. 

\item[{[20]}] R. V. Gavai 2000, Pramana J. Phys. {\bf 54}, 487. 

\item[{[21]}] C. Amsler et al. 2008, Phys. Lett. {\bf B67}, 1. 

\item[{[22]}] A. S. Kronfeld 2007, {\it Lattice gauge theory with
staggered fermions: how, where and why (not)}, 
PoS LAT2007:016,2007, arXiv:0711.0699 [hep-lat]. 

\item[{[23]}] J.J. Dudek, R.G. Edwards, N. Mathur, D.G. Richards 2008, Phys.Rev. {\bf D77}, 
034501. 

\item[{[24]}] Q. Mason et al. 2005 (HPQCD collaboration), Phys. Rev. Lett. {\bf 95}, 
052002. 

\item[{[25]}] I. F. Allison et al.2005, Phys. Rev. Lett. {\bf 94}, 172001.\\ A. Abulencia et 
al. 2006, {\it ibid.}, {\bf 96}, 082002. 

\item[{[26]}] R. V. Gavai 2006, Pramana J. Phys. {\bf 67}, 885. 

\item[{[27]}] See e.g. S. Aoki 2006, Int. J. Mod. Phys. {\bf A21}, 682. 

\item[{[28]}] See e.g. P. Jacobs and X.-N. Wang 2005, Prog. Part
Nucl. Phys. {\bf 54}, 443. 

\item[{[29]}] R. V. Gavai 2000, Pramana J. Phys. {\bf 55}, 15. 

\item[{[30]}] R.S. Bhalerao, J.-P. Blaizot, N. Borghini,
J.-Y. Ollitrault 2005, Phys. Lett {\bf B627}, 49. 

\item[{[31]}] D. P. Roy 2000, Pramana J. Phys. {\bf 54}, 3.\\ 
S. Uma Sankar, {\it ibid.}, 27.\\ 
A. Raychaudhuri, {\it ibid.}, 35.\\ 
A. S. Joshipura, {\it ibid.}, 119.\\ 
A. Bandyopadhyay, S. Choubey and S. Goswami 2005, Nucl. Phys. Proc. Suppl. 
{\bf 143}, 121.

\item[{[32]}] M. Sajjad Athar et al. 2006 (INO collaboration), 
{\it India-based Neutrino Observatory: project report}, Vol. 1, INO-2006-01.\\
B. Das, {\it Quest to gauge neutrino mass}: 
http://www.nature.com/nindia/2008/080831/full/nindia.2008.267.html.\\
D. Indumathi 2007, Mod. Phys. Lett. {\bf A22}, 2153.\\
D. Indumathi and M. V. N. Murthy 2005, Phys. Rev. {\bf D71}, 013001.\\
S. Choubey and P. Roy, {\it ibid.} {\bf D73}, 013006.

\item[{[33]}] See e.g. R. N. Mohapatra and P. B. Pal 2004, 
{\it Massive neutrinos in Physics and 
Astrophysics}, 3rd edition, World Scientific, Singapore. 

\item[{[34]}] U. Sarkar 2000, Pramana J. Phys. {\bf 54}, 101.

\item[{[35]}] P. Roy and O. Shanker 1984, Phys. Rev. Lett.{\bf 52}, 713. 

\item[{[36]}] V. K. B. Kota and U. Sarkar (eds.) 2008, {\it
Neutrinoless double beta decay}, Narosa, New Delhi. 

\item[{[37]}] F. T. Avignone, S. R. Elliott and J. Engel 2007, 
{\it Double beta decay, Majorana neutrinos and neutrino mass}, 
arXiv:0708.1033 [nucl-ex]. 

\item[{[38]}] R. G. Pillay 2007, in {\it Proc. Workshop on
neutrinoless double beta decay}, ed. V. Nanal, 
published by Tata Institute of Fundamental Research. 

\item[{[39]}] M. Raidal et al. 2008, {\it Flavour physics of leptons 
and dipole moments}, arXiv:0801.1826 [hep-ph]. 

\item[{[40a]}] M. Markevitch, 2005, {\it Chandra observation of 
the most interesting cluster in the universe},
arXiv:astro-ph/0511345.

\item[{[40b]}] See the MACHO webpage http://wwwmacho.anu.edu.au.

\item[{[41]}] R. M. Godbole and A. Gurtu (eds.) 2007, 
{\it Linear collider proceedings, international workshop, 
LCWS06} Bangalore, India, Pramana J. Phys. {\bf 69}, 777. 

\item[{[42]}] See e.g. D. Cline 2003, in {\it Tegernsee, Beyond the desert}, p587, 
astro-ph/0310439.

\item[{[43]}]  See e.g. S. Weinberg 1976, Phys. Rev  {\bf D13},
974. 1979, {\it ibid.}, {\bf D19}, 1207. 

\item[{[44]}] See e.g. M. Drees, R. M. Godbole and P. Roy 2004, 
{\it Theory and phenomenology of sparticles}, 
World Scientific, Singapore.\\ 
H. Baer and X. Tata 2006 {\it Weak scale supersymmetry}, 
Cambridge University Press, Cambridge, U.K.

\item[{[45]}] See e.g. R. Kaul and P. Majumdar 1981, 
Nucl. Phys. {\bf B199}, 36.\\ 
R. Kaul 1982, Phys. Lett. {\bf B109}, 19. 

\item[{[46]}] G. Bhattacharyya 2004, in {\it Frontiers in high energy 
physics}, Vol. 4, guest editors A. Raychaudhuri and P. Mitra, 
Allied Publishers, New Delhi, hep-ph/0108267. 

\item[{[47]}] J. C. Pati and A. Salam 1973, Phys. Rev. {\bf D8}, 1240. 

\item[{[48]}] H. Georgi and S. L. Glashow 1975, Phys. Rev. Lett, {\bf 32}, 438. 

\item[{[49]}] H. Georgi 1975, {\it Particles and fields}, 
Proc. APS Div Particles and Fields, (ed. C. Carlson), p575.\\ 
H. Fritzsch and P. Minkowski 1975, Ann. Phys. {\bf 93}, 193. 

\item[{[50]}] e.g. C. S. Aulakh, B. Bajc, A. Melfo, A. Rasin and 
G. Senjanovic 2001, Nucl. Phys. {\bf B597}, 89. 

\item[{[51]}] S. Raychaudhuri 2000, Pramana J. Phys. {\bf 55}, 171.\\ 
M. Schmaltz and D. Tucker-Smith 2005, Ann. Rev. Nucl. Part. Sci. 
{\bf 55}, 229.\\ 
M. Perelstein 2007, Prog. Part. Nucl. Phys. {\bf 58}, 247.\\ 
G. Bhattacharyya 2008, {\it Electroweak symmetry breaking and 
BSM physics (a review)}, arXiv:0807.3883.

\item[{[52]}] e.g. R. Sundrum 2004 {\it To the fifth dimension and
back} in ``TASI lectures: Boulder 2004; Physics in d = 4", 
hep-th/0508134.\\ 
R. Rattazzi 2003 {\it Cargese lectures on extra dimensions} 
in ``Cargese 2003: Particle physics and cosmology'', hep-ph/0607055.

\item[{[53]}] See e.g. S. B. Giddings and S. Thomas 2002, Phys. 
Rev. {\bf D 65}, 056010.

\item[{[54]}] See e.g. G. Landsberg 2006 J. Phys. G {\bf 32}, R337.

\item[{[55]}] See e.g. R. M. Godbole, S. K. Rai and S. Raychaudhuri
2007, Eur. Phys. J {\bf C50}, 979.

\item[{[56]}] U. Mahanta 2001, Phys. Rev. {\bf D 63}, 076006.\\ 
M. Chaichian, A. Datta, K. Huitu and Z-h. Yu 2002, Phys. Lett {\bf
B524}, 161.\\
P. K. Das, S. K. Rai and S. Raychaudhuri 2005, Phys. Lett. {\bf B618}, 221.

\item[{[57]}] G. Veneziano 1968, Nuovo Cim. A57, 190.\\
Y. Nambu 1969,
{\it Quark model and the factorization of the Veneziano amplitude},
in Proceedings, Conference On Symmetries, Detroit 1969.\\
L. Susskind 1970, Nuovo Cim. A69, 457.

\item[{[58]}] For details of  string quantisation, see:\\
M.B.  Green, J.H. Schwarz and E. Witten 1987, {\it Superstring Theory, Vols. 1,2},
Cambridge University Press, UK.\\
J. Polchinski, 1998, {\it String Theory, Vols. 1,2},
Cambridge University Press, UK.\\
B. Zwiebach 2004, {\it A first course in string theory},
Cambridge University Press, UK.

\item[{[59]}] P. Ramond 1971, Phys. Rev. {\bf D3}, 2415.\\
A. Neveu and J.H. Schwartz 1971, Nucl. Phys. {\bf B31}, 86.
 
\item[{[60]}] For a description of the central issues in Kaluza-Klein
compactification at the time, see E. Witten 1985, {\it Fermion Quantum
Numbers In Kaluza-Klein Theory}, 
Lecture given at Shelter Island II Conference, Shelter Island, N.Y., June 1983.

\item[{[61]}] M.B. Green and J.H. Schwarz 1984, Phys. Lett. {\bf
B149}, 117.

\item[{[62]}] For an exhaustive list of string theory papers in the first few years
after 1984, see the first reference in [58].

\item[{[63]}] D. J. Gross, J. A. Harvey, E. J. Martinec, R. Rohm 1984,
Phys. Rev. Lett. {\bf 54}, 502.

\item[{[64]}] P. Candelas, G. Horowitz, A. Strominger and E. Witten
1985, Nucl. Phys. {\bf B258}, 46.

\item[{[65]}] A. Sen 1985, Phys. Rev. Lett. {\bf 55}, 1846.\\
For a review of conformal invariance in string theory, see S.R. Wadia 1985,
{\it String theory and conformal invariance: A review of selected topics},
in proceedings of Superstrings, Supergravity and Unified Theories, 
ICTP High Energy Physics and Cosmology Workshop, Trieste, Italy, 1985.

\item[{[66]}] L. Alvarez-Gaum\'e, D. Z. Freedman and S. Mukhi 1981,
Ann. Phys. {\bf 134}, 85.\\ S. Mukhi 1986, {\it Nonlinear
$\sigma$-models, scale invariance and string theories: A pedagogical
review}, in proceedings of the TIFR Winter School in Theoretical High
Energy Physics, Panchgani, India, eds. V. Singh
and S.R. Wadia.

\item[{[67]}] See e.g. A. Dabholkar and C. Hull 2006, JHEP {\bf
0605}:009, and references therein.

\item[{[68]}] S. Kachru, R. Kallosh, A. Linde and S. Trivedi 2003,
Phys.Rev. {\bf D68}, 046005.

\item[{[69]}] S. Ashok and M. Douglas 2004, JHEP {\bf 0401}:060.

\item[{[70]}] L. Susskind 2005, {\it The Cosmic Landscape: String
Theory and the Illusion of Intelligent Design},
Little, Brown and Company.

\item [{[71]}] G. Horowitz and A. Strominger 1991, Nucl.Phys. {\bf
B360}, 197.\\
For early work see e.g. M. Duff 1988, 
Class. Quant. Grav. {\bf 5}, 189.

\item[{[72]}] J. Polchinski 1995, Phys. Rev. Lett. {\bf 75}, 4724.

\item[{[73]}] For a review, see A. Sen 1998, {\it An introduction to 
nonperturbative string theory}, in Cambridge 1997, ``Duality and
supersymmetric theories'', 297-413, e-Print: hep-th/9802051.

\item[{[74]}] J. Bagger and N. Lambert 2008, Phys. Rev. {\bf D77}, 065008.\\
S. Mukhi and C. Papageorgakis 2008, JHEP {\bf 05}:085.\\
O. Aharony, O. Bergman, D. Jafferis and J. Maldacena 2008, JHEP {\bf 10}:091.\\
For a comprehensive review of M-theory, see
K. Becker, M. Becker and J.H. Schwarz 2007, 
{\it String theory and M-theory: A modern introduction},
Cambridge University Press, UK.

\item[{[75]}] A. Strominger and C. Vafa 1996, Phys.Lett. {\bf B379}, 99.

\item[{[76]}] A. Sen 1995, Mod. Phys. Lett. {\bf A10}, 2081.

\item[{[77]}]  A. Dhar, G. Mandal and S. R. Wadia 1996,
Phys. Lett. {\bf B388} 51.\\
S.R. Das and S.D. Mathur 1996, Nucl. Phys. {\bf B478} 561.

\item[{[78]}] J. Maldacena 1998, Adv. Theor. Math. Phys. {\bf 2}, 231.

\item[{[79]}] E. Witten 1998, Adv. Theor. Math. Phys. {\bf 2}, 505.

\item[{[80]}] G. Policastro, D.T. Son and A.O. Starinets 2002,
JHEP {\bf 0209}, 043.\\
S. Bhattacharyya, V.E. Hubeny, S. Minwalla and M. Rangamani 2008,
JHEP {\bf 0802}:045.

\item[{[81]}] R.S. Bhalerao and R.V. Gavai, {\it Heavy ions at LHC: A
quest for quark-gluon plasma}, arXiv:0812.1619 [hep-ph]. 

\end{itemize}
 
\end{document}